 \documentclass[twocol]{ametsocV6.1}

\title{An extreme value method to study decadal hurricane wind trends}

\authors{%
Alexandre Payez,\aff{a}%
\correspondingauthor{Alexandre Payez, alexandre.payez@knmi.nl}
Ad Stoffelen,\aff{a}
Cees de Valk,\aff{a}
and Rianne Giesen\aff{a}
}

\affiliation{\aff{a}{Royal Netherlands Meteorological Institute (KNMI), De Bilt, The Netherlands}}

\abstract{%
This paper presents a method developed using techniques from extreme value theory to estimate smooth wind-speed percentiles, allowing us to consider more extreme wind speeds while being less sensitive to the noise that stems from the scarcity of extreme data.
A reliable characterisation of wind extremes is the first required step for studying decadal trends in tropical-cyclone and extra-tropical-cyclone winds.
We develop a percentile-smoothing method using ASCAT-A Level-3 products, focusing on a number of tropical basins (Caribbean and Atlantic), estimate the uncertainty with the block-bootstrap technique to address the issue of dependency, and apply our method to both scatterometer winds (ASCAT-A at two different resolutions) and collocated ERA5 model data.
The results obtained are very robust at basin level, without having to rely on a strong assumption for the distribution tail: they are very consistent whether we use exponential fits, generalised-Pareto fits, or even no fit at all, down to at least truly extreme wind percentiles such as 99.999\textsuperscript{th} (main result), and remain quite consistent within uncertainties down to 99.9999\textsuperscript{th}.
As ensuring scientifically sound decadal-trend conclusions would require going back sufficiently in time, spanning the lifetimes of different instruments with different characteristics and extreme-wind statistics, a natural follow-on study would be to apply this method not only to ASCAT, but also to its predecessors on QuikSCAT and ERS\hspace{1pt}---\hspace{1pt}comparing each scatterometer individually against ERA5, and also to each other as partial overlaps exist between instruments.
}

\begin{document}

\maketitle

%
%
%
\statement

Global warming is known for intensifying various types of extreme weather such as heat waves and extreme precipitation, compared to the pre-industrial era.
Quantifying and providing conclusive evidence of whether, how, and to what extent climate change impacts hurricane-force winds over the ocean is a far more difficult question to answer however.
Properly assessing the presence of geophysical trends will require considering time series of extreme winds over multiple decades, carefully comparing data from different instruments --- knowing that the first mandatory step is of course to develop a robust method able to reliably identify changes in extreme winds.
This work provides such a method.

\section{Introduction}

Weather and climate extremes are duly receiving a lot of attention, with extreme storms such as tropical cyclones (also called hurricanes or typhoons) chiefly among them.
This is  evidenced by numerous studies and series of regularly updated reports about extreme storms, not least from the World Meteorological Organization (WMO), both in terms of observations~\citep{Knutson_etal:2019} and model projections~\citep{Knutson_etal:2020}.
The question of whether and how various extremes might be affected by climate change is obviously an important and recurrent topic in the Intergovernmental Panel on Climate Change assessment reports as well~\citep{IPCC_Extremes:AR6}, and has in fact sometimes even been the subject of a special stand-alone IPCC report~\citep{IPCC_SREX_Chap3:2012}.
The importance of extremes is also well recognised by international organisations such as the European Space Agency (ESA), which recently placed `extremes and hazards' at the heart of its Earth Observation Science Strategy 2040, as one of the overarching themes~\citep{ESA:2024}.

For extreme storms specifically, the question of detecting decadal trends as well as the relation to climate change is known to be evidently difficult however \citep{Coumou_Rahmstorf:2012}, and has led to a number of controversial claims. This is also evident from the WMO assessment done by \citet{Knutson_etal:2019}.
Owing e.g.\@ to their very localised nature (both in time and in space), destructive power, complexity which cannot be captured with a single metric, and the large natural variability in the occurrence of such events, the study of extreme storms comes with additional challenges compared to some other extremes.

Atmospheric reanalyses, such as ERA5~\citep{Hersbach:2020} which is produced by the European Centre for Medium-Range Weather Forecasts (ECMWF), are a main vehicle to study climate trends.
To help detect trends faithfully, one version of an atmospheric model or an Earth-system model may be used to analyse the weather over many decades.
However, since the main input to such analyses are global observations, one should note that the distribution of global observations did change fundamentally over past decades.
The model is the same, but the nature of the inputs and their quality have been changing.
The main input to reanalyses over the last century are in-situ observations, often surface-based, whereas over this century satellite observations, mainly upper air, are the main contributor to the quality of reanalyses.
It is yet unclear whether such changes in the observation input to the reanalyses over past decades affect the properties of modelled phenomena such as tropical cyclones; hence, assessments of trends in the occurrence and intensity of tropical hurricanes are hampered by this.

Long high-quality observation records may arguably be better suited to detect climate trends 
\citep{Verhoef_etal:2017}.
For an up-to-date overview of past and present satellite observations in tropical cyclones, the interested reader is referred to \citet{Velden_etal:2025}.
We should note however that not many of these measurement techniques have been employed over several decades with a constant quality and a high enough coverage to derive extreme wind climate trends.
Significantly longer climate-quality time-series of measurements, involving both past and future instruments, will be needed before the presence of a detectable decadal-trend signal in hurricane wind speeds may be assessed with high confidence, based on solid scientific evidence.

In this study, undertaken in the context of the MAXSS project of the European Space Agency, to help investigate changes in extreme wind speeds associated with tropical hurricanes and similar storms over the ocean, we use surface winds obtained with the ASCAT scatterometer (MetOp satellites~\citep{EUMETSAT_EPS:2007}).
As our analysis is based on percentiles of ASCAT wind, it is little influenced by the in-situ reference used to calibrate the observations:
using another, or more consolidated, wind-speed reference
would mainly associate each percentile with another wind speed, but not dramatically change the trend detection procedure results.

Scatterometer winds are also well suited for this study. Indeed, the quality of hurricane winds from Synthetic Aperture Radar (SAR) observations, wind scatterometers observations, and operational winds from ECMWF were investigated by \citet{Ni_etal:2022a}.
Taking into account the differences in calibration and spatial resolution, and using a triple collocation method, they confirmed the high quality of ASCAT scatterometer winds in tropical hurricanes, finding that ASCAT wind speeds have a random error of about 15$\%$.
\citet{Mayers_Ruf:2020} also previously discussed the topic of converting TC intensity estimates defined at different scales.

The paper is organised as follows. For studying trends in extreme winds, we elaborate first on spatio-temporal aspects and extreme-percentile smoothing. Subsequently, the datasets used to develop the method are discussed. Extreme value theory concepts are furthermore presented and sample fits at Caribbean basin level are evaluated.
Following a discussion and summary, an outlook is provided.

\section{Studying trends in extreme winds}
\label{sec:extremewindtrends}

\subsection{Spatio-temporal aspects}

Generally speaking, the context of this work is the identification and study of potential statistically significant changes in extreme wind speeds over the global ocean, from tropical cyclones (TC) to extra-tropical cyclones (ETC) and polar lows (PL). To this aim, scatterometers, which have now been delivering surface wind data from radar back-scatter measurements over multiple decades, are particularly well-suited: in particular, they present a clear advantage over altimeters, as scatterometers have a much broader swath.\footnote{The Seawinds scatterometer on board of QuikSCAT, which had a particularly broad coverage, gives a striking demonstration of this as it was in fact able to cover about 90\% of the world's oceans in 24 hours; see e.g.\@ \citet{OSI-SAF_ATBD}.}

While scatterometers do provide a good spatial coverage of 10-m stress-equivalent surface wind speeds $U_{10s}$ over the globe, for a trend analysis we should also worry about the temporal dimension.
In particular, for a decadal trend analysis of extreme wind speeds, we should be wary that if our data do not cover a long enough period, the trend estimates would likely be sensitive to decadal-scale climate patterns like the El Niño Southern Oscillation (ENSO), and it could in fact be very difficult to isolate ENSO effects from the general trend and to say anything really meaningful about decadal changes. 
To obtain sufficient evidence for extreme wind-speed decadal trends, one ought to notably go further back in time and consider also space-borne scatterometer instruments preceding ASCAT, and notably include Seawinds on QuikSCAT and SCAT on ERS.

\subsection{Further scientific requirements for discussing trends}

In order to meaningfully discuss trends, one important requirement that should not be overlooked is that the data considered should be of similar quality and similar sampling. When it is not the case, it may not be possible to extract useful scientific information. 

Different instruments have different characteristics and different extreme wind statistics that can notably be linked to calibration and rain contamination.
For that reason, the approach that we advocate is not to aggregate time series of different instruments into one record but instead to compare each of them separately against a common reference dataset such as ERA5.

\subsection{Studying extreme events in terms of extreme percentiles}

As was already understood in a number of studies on extreme winds done at the Royal Netherlands Meteorological Institute (KNMI) in recent years \citep{Marseille_etal:2019, Giesen_Stoffelen_OSR6:2022}, percentiles are particularly well suited to study extremes if we want to consider probabilities as fractions of time, and not annual probabilities of exceedance as is more common in extreme value analysis.
They are indeed much easier to estimate, and just as useful for the answering the question at hand.
Interestingly, it should remain possible to a posteriori adjust the wind values of the percentiles obtained by using an different Geophysical Model Function (GMF).
This is notably relevant since the use of either dropsondes or buoys is known to give rise to differences in calibration \citep{CHEFS:2020}\hspace{1pt}---\hspace{1pt}buoys being in good agreement with ECMWF, and dropsondes being the historical standard used and relied upon by the operational hurricane community. 

Both \citet{Polverari_etal:2022} and \citet{Ni_etal:2022a} discuss the issue of in-situ calibration at high and extreme winds, where dropsondes are the main wind calibration reference. Dropsondes are used to calibrate airplane instruments, such as SFMR, but also satellite surface winds from radiometers and scatterometers. These observations are moreover used to tune atmospheric model winds to comply with these observations. For winds globally occurring up to 25 m/s, well-calibrated winds from moored buoys are used to calibrate satellite winds and tune model winds, while these are inconsistent with dropsonde winds \citep{stoffelen2021} in the common occurrence range of 20 to 25 m/s, up to 40$\%$.
While \citet{Rabaneda_etal:2025} do not solve this wind speed reference inconsistency, they at least aim to provide a consistent wind reference for hurricane advisory centres for different airplane and satellite extreme-wind measurements; see Table~\ref{table:adjustement_between_buoys_and_SFMR_scale}.

\begin{table}
    \centering
    \begin{tabular}{c|c}
        Original $U_{10s}$ winds & Adjusted $U^*_{10s}$ winds
        \\(buoys) & (SFMR/dropsondes)
        \\[.2cm]
        \hline
        20 m/s &25.1 m/s\\
        22 m/s &28.7 m/s\\
        24 m/s &32.5 m/s\\
        26 m/s &36.4 m/s\\
        28 m/s &40.5 m/s\\
        30 m/s &44.7 m/s\\
        32 m/s &49.0 m/s\\
        34 m/s &53.6 m/s\\
        36 m/s &58.2 m/s\\
        38 m/s &63.0 m/s\\
        40 m/s &68.0 m/s
    \end{tabular}
    \caption{
    Table showing a few values obtained using the wind-speed adjustment formula recently derived in \cite{Rabaneda_etal:2025} 
    to link the original ASCAT winds calibrated against buoys $U_{10s}$, to ASCAT winds adjusted to the SFMR scale (dropsondes) $U^*_{10s}$.
    In this work, we will stick to the original ASCAT stress-equivalent wind-speed values $U_{10s}$ throughout.
    }
    \label{table:adjustement_between_buoys_and_SFMR_scale}
\end{table}

\subsection{Extreme-percentile smoothing}
\label{sec:extremepercentileinterp}

Extremes are rare by definition. This implies that working with e.g.\@ an empirical 99$^{\rm th}$ percentile of wind speed can obviously be very noisy, since most of the available wind data is thrown away, leaving only the highest 1\% wind values.
For that reason, the idea pursued in the present work is to exploit extreme value theory techniques. We are therefore interested in percentile-smoothing techniques, which should then give us more precise results. It will moreover allow us to consider even higher percentiles, i.e.\@ more extreme winds, and make sure to better focus the analysis on tropical cyclones. The idea pursued in the present work is to exploit extreme value theory techniques for this purpose.

\section{Datasets used}

To develop our method, we use ASCAT Level-3 wind-speed data and focus on a number of tropical basins, namely the Caribbean, the North Atlantic in the 0--30°~N latitude range, and the South Atlantic in the 0--30°~S latitude range.
Advantages of using ASCAT scatterometer winds are that long time-series are available, and that it is very stable over time. They were also used in an earlier extreme-wind study based on percentiles~\citep{Giesen_Stoffelen_OSR6:2022} which appeared in the Sixth Copernicus Marine Service Ocean State Report (OSR6)\hspace{1pt}---\hspace{1pt}however that study was limited to 99\textsuperscript{th}-percentile wind speeds, which are actually too low to correspond to extreme storms such as TCs.

We use ASCAT L3 reprocessed ocean surface wind datasets for the period 2007-2021 distributed by the Copernicus Marine Service. As MetOp-A retired in November 2021, datasets from the equivalent MetOp-B ASCAT scatterometer were used for the year 2021. We did not apply any masking to the data, to improve on what was done in the OSR6 study~\citep{Giesen_Stoffelen_OSR6:2022} which did not consider grid cells with certain quality flags set (e.g.\@ related to wind direction). Their masking method removed part of the highest observed wind speeds in the datasets, which we prefer to include in this study.

\bigskip

\begin{figure}[h]
    \centerline{\includegraphics[width=19pc, angle=0, trim=0 0 0 0, clip]{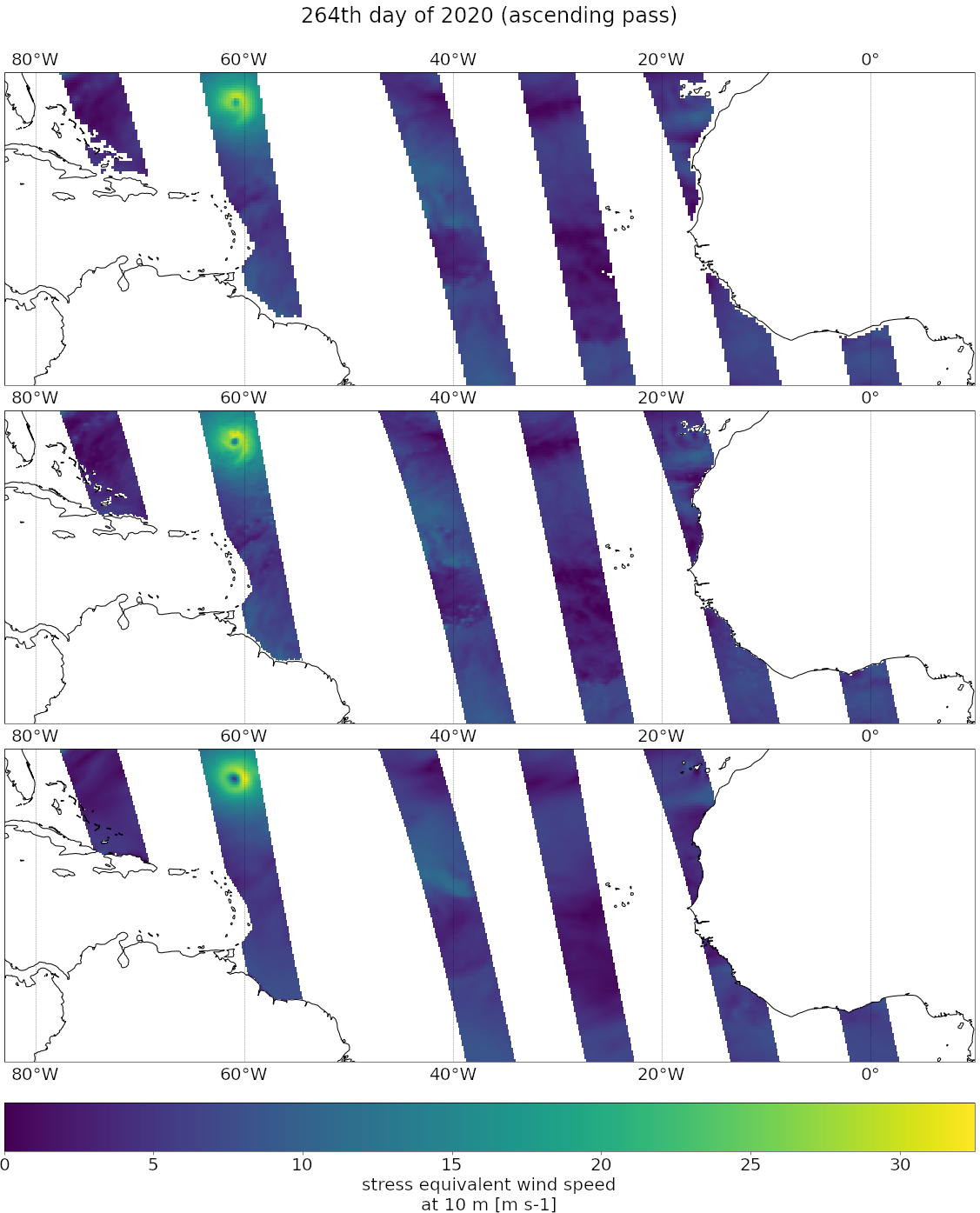}}
    \caption{ASCAT-A 0.25° (\emph{top}) and ASCAT-A 0.125° (\emph{middle}) L3 products, North Atlantic (0--30° N): example of a day with extreme winds in 2020 (September 20th). As expected, larger $U_{10s}$ wind speeds ($> 30$~m/s) are reached in the higher-resolution product, which moreover displays sharper gradients. The collocated ERA5 stress-equivalent model winds from the 0.125° L3 products are also shown (\emph{bottom}).}   \label{fig:exploratory_data_analysis_sample_map_atlantic_basin}
\end{figure}

\noindent{}For this work, we shall consider ASCAT-A L3 products with two different resolutions:

\paragraph{025}

\begin{itemize}
    \item L3 product from the Copernicus Marine Service based on 25-km swath-aligned Level-2 (L2) data (50~km resolution), regridded on a 0.25° (1 degree $\sim$110~km at Equator) regular latitude, longitude grid.
    \item lower noise because more data are accumulated.
\end{itemize}

\paragraph{012}

\begin{itemize}
    \item L3 product from the Copernicus Marine Service based on 12.5-km grid L2 data (25 km resolution), regridded on a 0.125° spacing grid.
    \item strong gradients are better represented but higher noise.
\end{itemize}

Figure~\ref{fig:exploratory_data_analysis_sample_map_atlantic_basin} shows an example of a day with extreme winds (2020 Hurricane Teddy).
To study tropical-cyclone winds, higher-resolution 012 data are particularly well suited, given their stronger gradients and wind speeds, seen in the central panel of Figure~\ref{fig:exploratory_data_analysis_sample_map_atlantic_basin}; see also e.g.\@ \citet{OSI-SAF_ATBD}. Nonetheless, using 025 is also interesting (top panel): not only for developing purposes and to compare against the final 012 results, but also for follow-on studies considering time series from scatterometers before the ASCAT period, as 25-km grid products also exist for both ERS SCAT and QuikSCAT Seawinds.
Moreover, together with the scatterometer data, the L3 datasets contain collocated ERA5 stress-equivalent model wind speeds, which are included for comparison purposes (bottom panel).

\section{Extreme value theory}

As noted in Section~\ref{sec:extremewindtrends}\ref{sec:extremepercentileinterp}, our interest is in percentiles: wind speeds not exceeded in a given percentage of time. 

Let $F_X$ be the cumulative distribution function (CDF) of the wind speed, so $F_X(x) = P(X \leq x)$ represents the mean fraction of time during which a value $x$ is not exceeded.
The wind-speed percentile for a percentage $N\%$ can thus be written as $F_X^{-1}(N/100)$.

Because for percentages close to 100 (i.e.\@ in the upper tail), empirical percentiles are inaccurate, we want to smooth these percentiles.

Extreme value theory \citep{Coles:2001, Haan_Ferreira:2006} presents a reasonable approach to smoothing of percentiles in the tail.
Applied to wind speed, it assumes that the difference between the wind speeds exceeded with probabilities $p$ and $p/\lambda$ (with $\lambda>0$), scaled by a suitable positive number depending on $p$, tends to a non-constant function of $\lambda$ when the probability $p$ tends to 0.
This assumption can be written as:
\begin{equation}
    \label{eq:GP}
    \frac{F_X^{-1}(1-p/\lambda)- F_X^{-1}(1-p)}{a(p)} \to h(\lambda)
\end{equation}
as $p \to 0$ for some positive function $a$. This is a very weak assumption.
It can be derived that the limiting function $h$,  which is independent of $p$, must be of the form
\begin{equation}
    \label{eq:h}
  h(\lambda)=  (\lambda^\gamma -1)/\gamma
\end{equation}
for some real number $\gamma$ (the right-hand side is to be read as $\log \lambda$ if $\gamma= 0$); see e.g.\@ \cite{Haan_Ferreira:2006}.

From the equations \eqref{eq:GP} and \eqref{eq:h}, all well-known distributions in extreme value theory can be derived \citep{Haan_Ferreira:2006}:
\begin{itemize}
    \item 
    the limiting distribution of the shifted and scaled maximum of $n$ independent and identically distributed random variables, when $n$ increases to infinity (or similarly, considering the maximum of a stationary stochastic process over a finite time-interval, when the length of the interval tends to infinity; see \citet{Leadbetter_Rootzen:1988});

    \item 
    the limiting Generalised Pareto (GP) distribution of the scaled excess above a high threshold, conditional on it being positive \citep{balkema1974residual}. This limit is exactly equivalent to the limit \eqref{eq:GP}, \eqref{eq:h}, so we will refer to \eqref{eq:GP}, \eqref{eq:h} as the Generalised Pareto approximation also.
    For a fixed value of $p$, $a(p)$ in Equation~\eqref{eq:GP} corresponds to the scale parameter of the Generalised Pareto distribution.
    A special case is the exponential distribution, which corresponds to \eqref{eq:GP}, \eqref{eq:h} with $\gamma= 0$.
    
    It is most frequently used in the peak-over-threshold method \citep{Leadbetter:1991}, which identifies clusters of values exceeding a high threshold and selects the highest peak from each cluster, and then estimates a Generalised Pareto approximation to the distribution of the excesses of these peaks above the threshold. Because these peaks are approximately independent, we can derive frequencies of exceedance of high values from this distribution.
\end{itemize}

We should stress that the basic regularity assumption  on the tail (Equations~\eqref{eq:GP} and \eqref{eq:h}) has nothing to do with dependence or independence. Therefore, it can be readily applied to the smoothing of percentiles of wind speed derived from scatterometer data, irrespective of spatial or temporal dependence in the data.

In fact, the only effect of dependence on  estimates of percentiles is that they are less precise than estimates from independent data, so in computing error variances or confidence intervals, we will need to account for that dependence somehow.

\section{Smoothing percentiles above a high threshold}
\label{sec:interp_percentile_method}

\begin{figure}[h]
 \centerline{\includegraphics[width=19pc, angle=0]{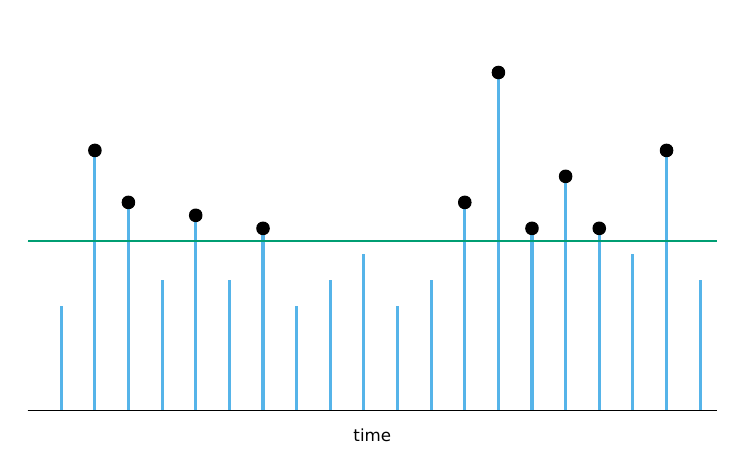}}
    \caption{Sketch representing the selection of values exceeding a threshold from a time-series.}
    \label{fig:blockmaxima_peakoverthreshold}
\end{figure}

Seeing that our objective is to study extreme winds in terms of percentiles, we propose to choose the threshold itself to be a specific high percentile of the considered dataset, say the $N_{\rm threshold}$-th percentile.
Then $x_{\rm threshold}= F_X^{-1}(N_{\rm threshold}/100)$ is the threshold for our analysis; see Figure \ref{fig:blockmaxima_peakoverthreshold}.

As usual, below the threshold, the data is considered sufficiently reliable such that no model needs to be introduced, while above the threshold the tail of the wind-speed distribution is to be fitted (e.g.\@ using exponential or generalised Pareto distributions).

\bigskip

The idea is then that percentiles can therefore be obtained as follows:
\begin{itemize}
    \item percentiles up to the $N_{\rm threshold}$-th percentile are simply calculated based on the data itself;
    \item beyond the threshold, percentiles higher than the $N_{\rm threshold}$-th percentile are calculated from the tail distribution fitted to the  data exceeding the threshold: in terms of probabilities instead of percentages, this approximation is given by Equations~\eqref{eq:GP} and~\eqref{eq:h}, with $p= 1-N_{\rm threshold}/100$ and $\lambda>1$.
    In Equation~\eqref{eq:GP}, $F_X^{-1}(1-p)$ is then the wind speed that corresponds to the threshold, while $F_X^{-1}(1-p/\lambda)$ is the wind speed that corresponds to an even higher percentile, beyond the threshold, for $\lambda>1$ (with a probability which is then $\lambda$ times lower).
\end{itemize}
Technically, the tail fit from the data above the threshold results in the conditional distribution of only the values exceeding the threshold. For values above the threshold $x_{\rm threshold}$, this conditional distribution needs to be converted to the unconditional tail distribution $F_X$ in \eqref{eq:GP}.

Taking a high enough percentile value as the threshold can assure us that we are only looking at TC winds; however, the higher the threshold, the less data available for the fit, meaning that there is a balance. As we do not want to compromise on either the need for the threshold to be high enough to be mostly sensitive to TC winds, nor on the need for preserving enough data in order to fit the tails, our study will actually be done at basin level.

Separately for each basin that we consider, we will bundle together all the wind-speed information that we have from all the available pixels for which we have data, and use the resulting sample as being representative from the population of wind speeds in that basin over the considered period (which could span one or multiple years). Our objective is then to derive the extreme percentiles of the entire wind-speed population in that basin over that period.
We will not apply our method at pixel level to produce maps of extreme percentiles winds, as there would be too little data available for reliable results at tropical-cyclone wind speeds.

\begin{figure}[ht]
    \centerline{\includegraphics[width=19pc, angle=0, trim = 0 0 0 0, clip]{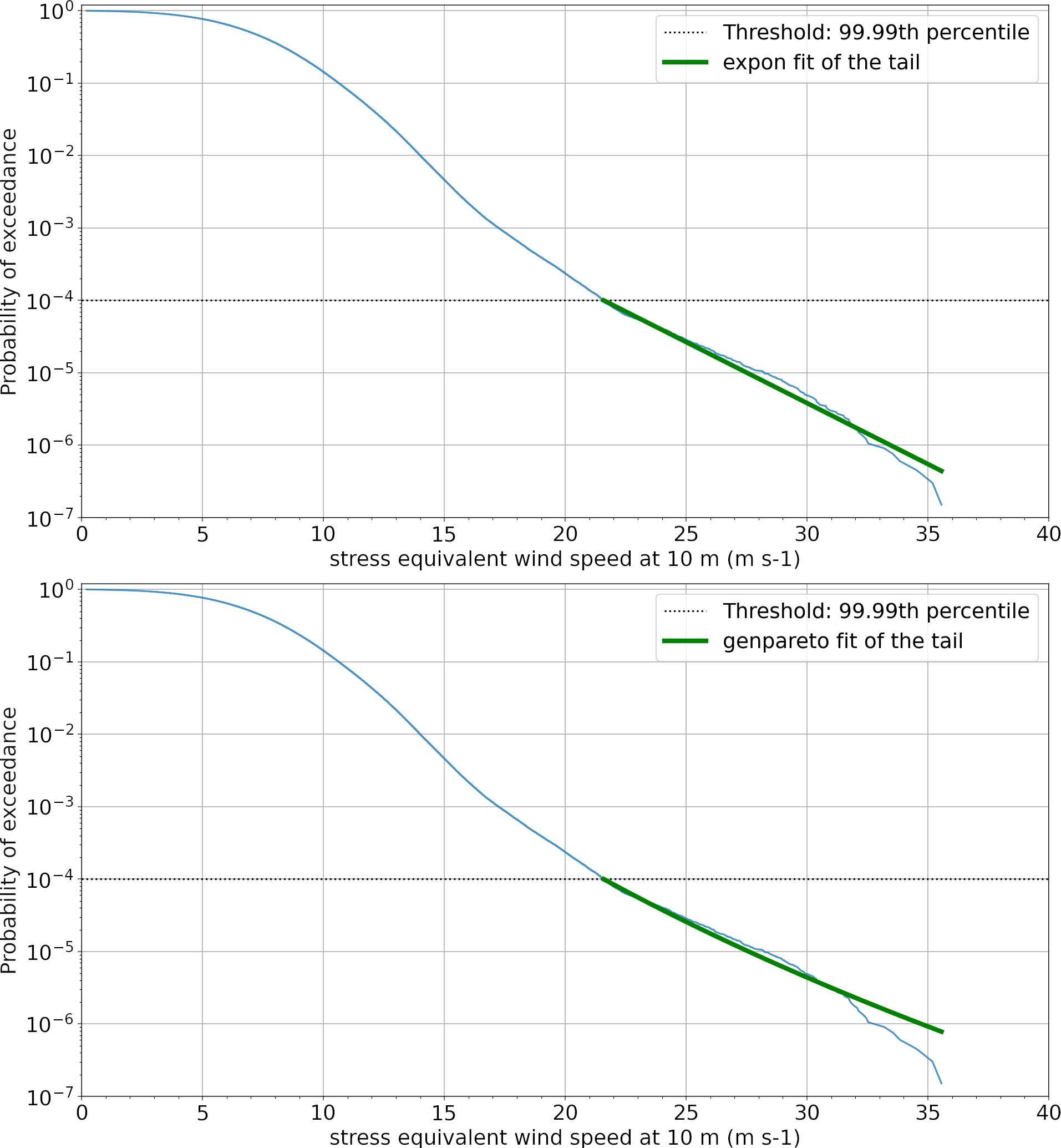}}
    \caption{Caribbean basin (ASCAT-A\_012): probability of exceedance for the year 2020. The blue curve is the empirical probability of exceedance, calculated from the empirical cumulative distribution function (ECDF) and Equation~\eqref{eq:link_exceedance_percentiles}. The green curves are the exponential (\emph{top}) and generalised-Pareto (\emph{bottom}) fits of the tail distribution\hspace{1pt}---\hspace{1pt}using the 99.99$^{\rm th}$ percentile of the basin wind-speed dataset as threshold.}
    \label{fig:Caribbean_basin_2020_proba_exceedance_fits}
\end{figure}

To study extremes, one usually considers the probability of exceedance $P(X>x)$, 
which relates to the $j$-th percentile $x_j$ via
\begin{equation}
    P(X>x_j) = 1-F_X(x_j) = 1 - \frac{j}{100}, \qquad j \in [0,100];
    \label{eq:link_exceedance_percentiles}
\end{equation}
meaning e.g.\@ that the 99.99\textsuperscript{th} percentile has a probability of exceedance of $10^{-4}$.
The probability of exceedance is usually plotted in log-scale as a function of the wind speed in linear scale; cf.\@ Figure~\ref{fig:Caribbean_basin_2020_proba_exceedance_fits}.

\section{Estimating uncertainty with bootstrapping}

\begin{figure}[h!!!]
    \centerline{\includegraphics[width=19pc, angle=0, trim = 0 0 0 0, clip]{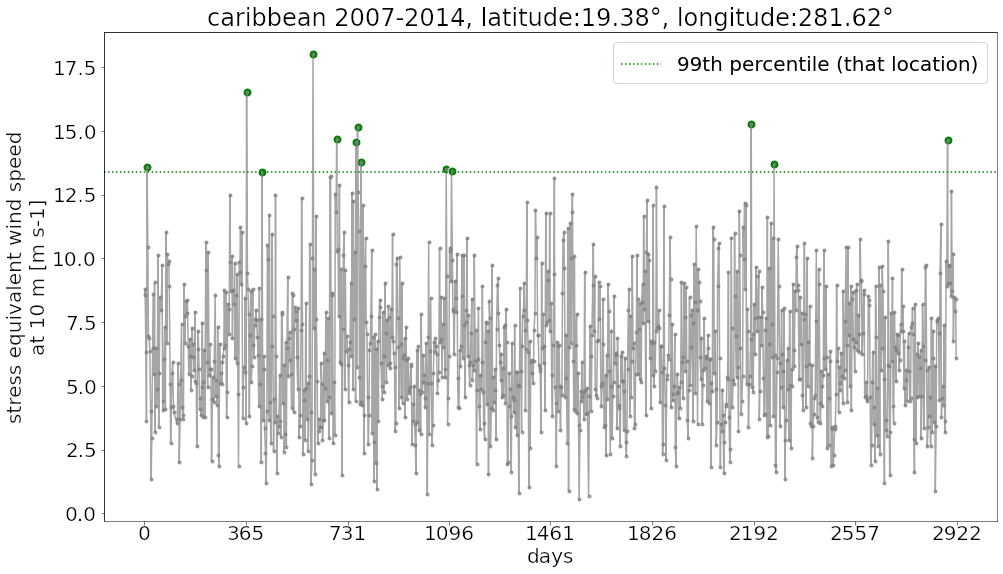}}
    \caption{Time series showing all the wind data for a given pixel location in the middle of the Caribbean basin, from January 2007 to December 2014 included (ASCAT-A L3 025 data; ascending passes only), with a 99th percentile threshold.
    This shows how, if working at pixel level, one would only preserve a handful of wind-speed measurements above a 99th-percentile threshold\hspace{1pt}---\hspace{1pt}even though the time series itself in this example does contain 8 years of scatterometer data.
    }
    \label{fig:PoT_timeseries_threshold_and_throwing_data_away}
\end{figure}

\begin{figure}
    \centerline{\includegraphics[width=14pc, angle=0]{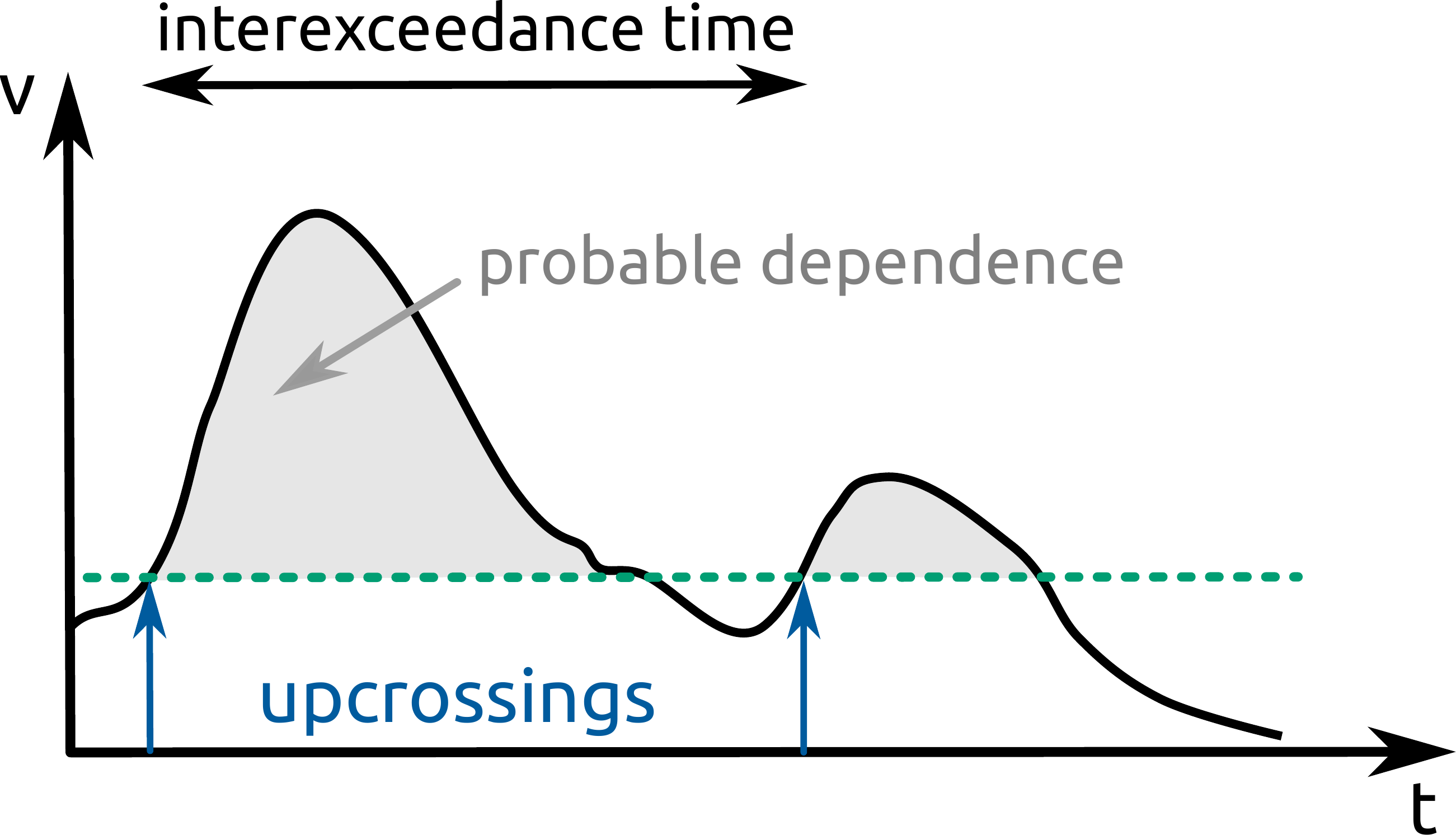}}
    \caption{Illustration of the issue of dependence as a function of time, in a peak-over-threshold context.}
    \label{fig:probable_dependence_sketch}
\end{figure}

When studying extremes, data is necessarily scarce (see e.g.\@ Figure~\ref{fig:PoT_timeseries_threshold_and_throwing_data_away} for a striking illustration).
Therefore, rather than sacrificing the data even more when attempting to ensure that they are independent (cf.\@ Figure~\ref{fig:probable_dependence_sketch}), the approach that we take is instead to provide an estimate of the variance/random error associated to our estimators for the large percentiles that we consider, using the bootstrap technique (block bootstrap) \citep{Efron:1979, Kunsch:1989}.

What is crucial to understand is that the objective with the bootstrap is not to remove the dependence in the data; the aim is instead to estimate this dependence, and therefore one needs instead to retain temporal and spatial correlations.
What is done is to start from the original data, and to each time generate a different dataset, which is like a brand new sample characterised by the same temporal and spatial correlations as the original one. Each random sample is such that part of the data will be omitted while part of it will be repeated.

\subsection{Resampling the data}

In practice, we actually resample the days in the period of interest. Whenever a specific day is randomly selected, all the corresponding swaths for that day are taken\hspace{1pt}---\hspace{1pt}i.e.\@ exactly data layers as shown in Figure~\ref{fig:exploratory_data_analysis_sample_map_atlantic_basin}, both for ascending and descending passes. Doing so ensures that the spatial correlations are preserved. On the other hand, preserving the temporal correlations that exist in the data is done by selecting consecutive days whenever we pick one at random. Seeing that the objective is to preserve timescales typical for the problem of interest, the timescale should be similar to the time it takes for a tropical cyclone to cross the basin, i.e.\@ about one week.

In addition, the days themselves are not drawn at random. Indeed, one moreover needs to be careful to preserve seasonality. 
The original sample is first split into the 12 calendar months, bundling days irrespectively of the year they are coming from (e.g.\@ all days from February 2019 and February 2020 would be put together as "February" for a sample that would correspond to the 2019--2020 period).
Then, when resampling the dataset, the number of days falling in each of the 12 calendar months from January to December will remain exactly the same as in the original sample (e.g.\@ 57 days for February in our 2019--2020 example). 
The point is to preserve the weight of each season in the sample, to be identical to what was in the original dataset.

\begin{figure}[h]
    \centerline{\includegraphics[width=16pc, angle=0, trim = 0 0 35 35, clip]{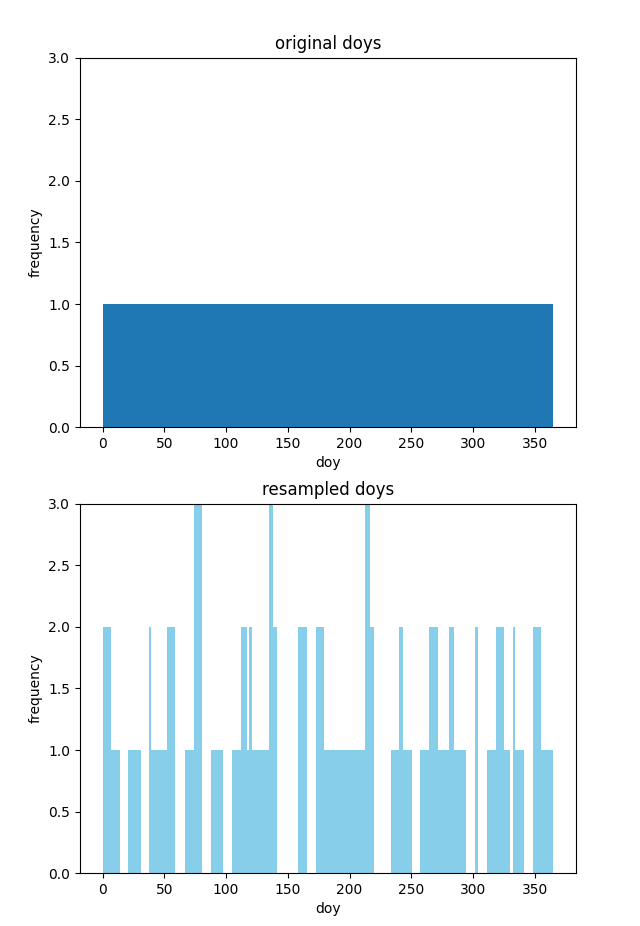}}
    \caption{Example realisation of resampled days with the block-bootstrap method (see text) for the Caribbean basin in the year 2020 (doys are number of days from January 1\textsuperscript{st}, starting at 0).}
    \label{fig:blockbootstrap_resamplingdays}
\end{figure}

\begin{figure}[h]
    \centerline{\includegraphics[width=19pc, angle=0, trim = 0 0 0 0, clip]{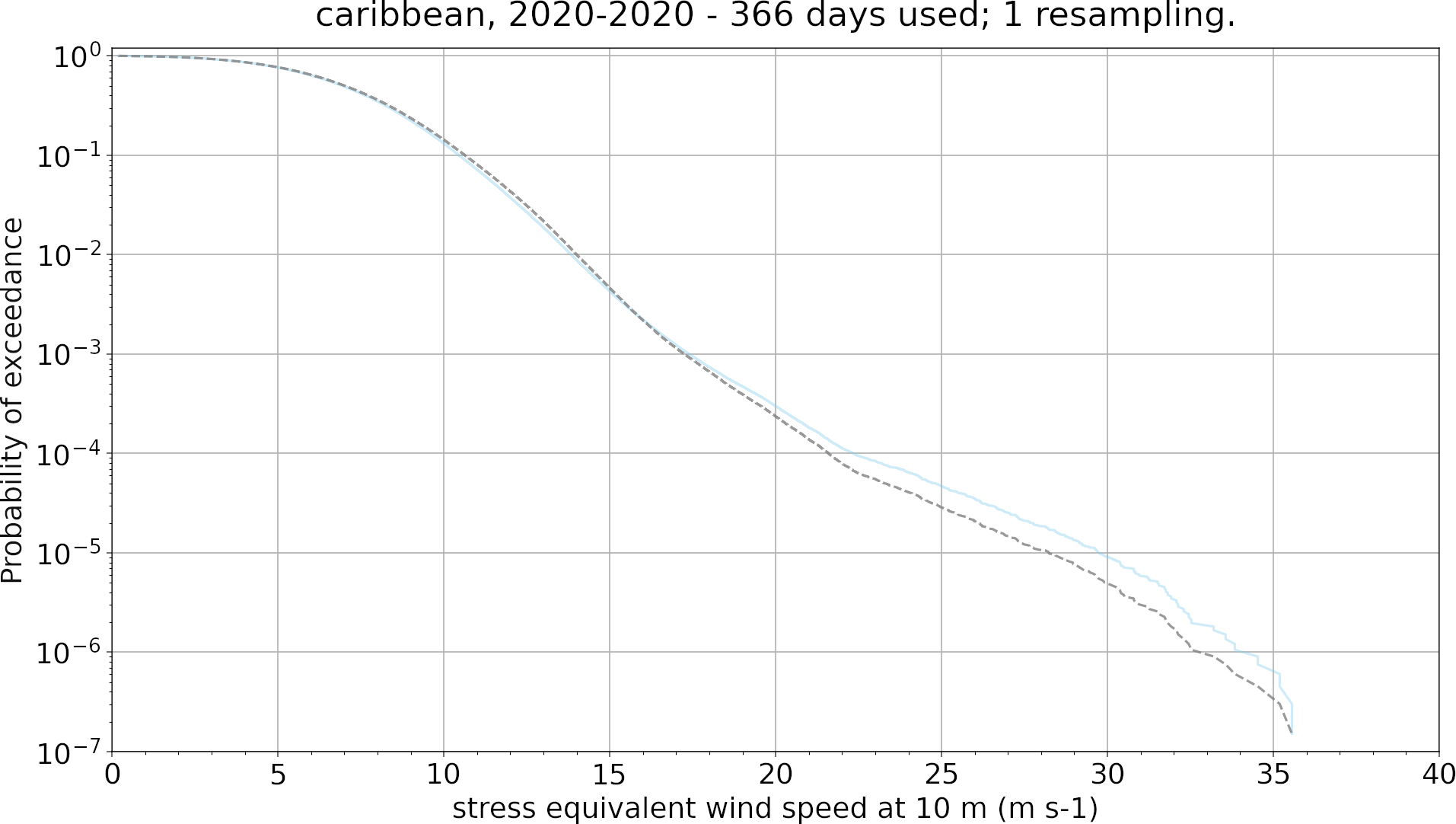}}
    \caption{ASCAT-A 012 data. Illustrating the empirical probability of exceedance calculated for one instance of block-bootstrap resampling of the Caribbean-basin wind speeds, for the year 2020. The original data is shown in dashed grey; the resampled data in blue.}
    \label{fig:caribbean_2020_012_blockbootstrap_1_resampling}
\end{figure}

One actual realisation of the temporal resampling done for this block-bootstrapping that preserves seasonality is shown in Figure~\ref{fig:blockbootstrap_resamplingdays}. Correspondingly, Figure~\ref{fig:caribbean_2020_012_blockbootstrap_1_resampling} shows the probability of exceedance in both the original and resampled data cases. 
In passing, let us point out that the extreme-wind content of that specific resampled dataset is larger than in the original sample.

\subsection{Multiple resampling and fitting}
\begin{figure}[t!!]
    \centerline{\includegraphics[width=19pc, angle=0, trim = 0 0 0 0, clip]{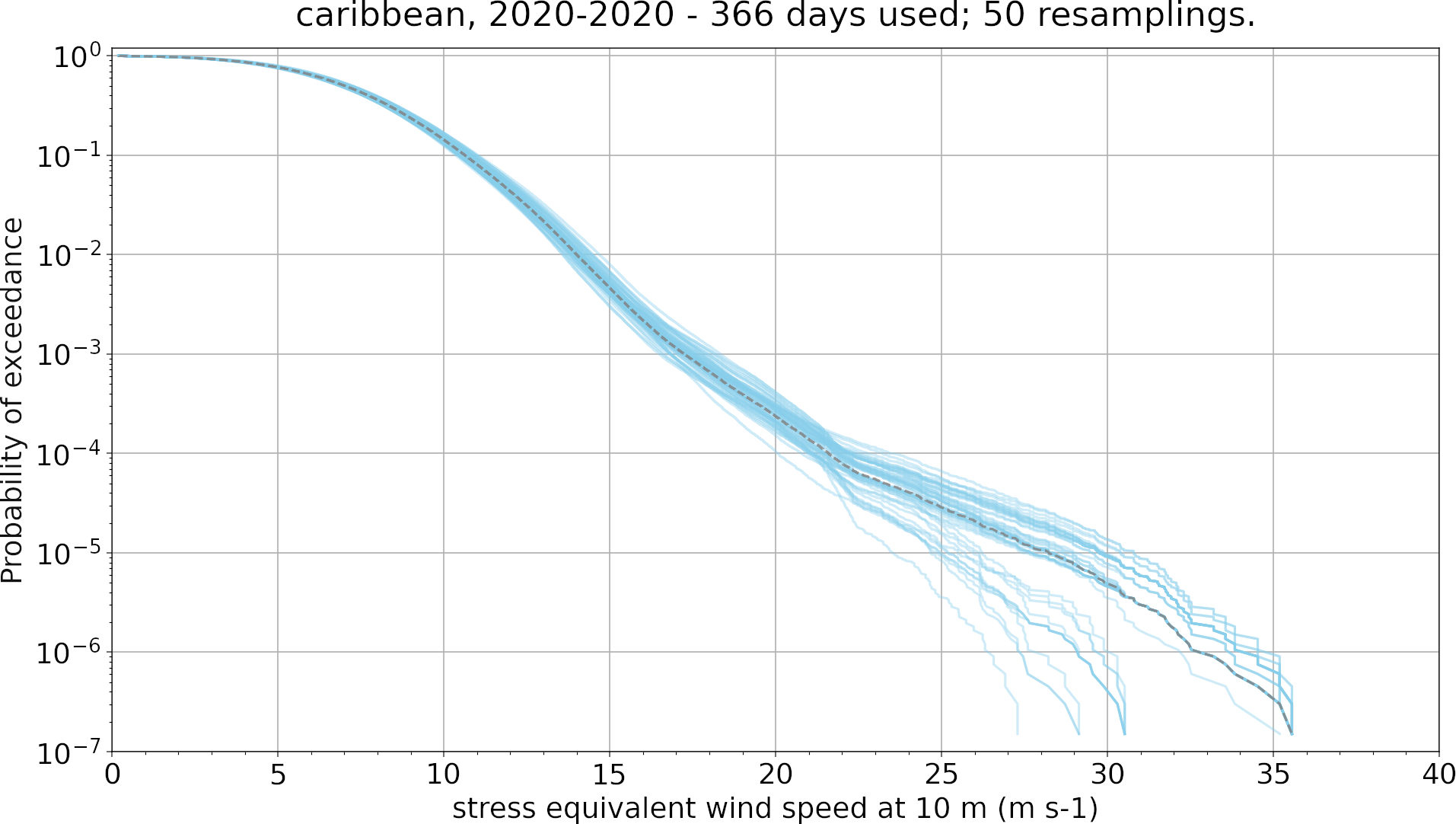}}
    \caption{ASCAT-A 012 data. Empirical probability of exceedance calculated for 50 block-bootstrap resamplings of the Caribbean-basin wind speeds, for the year 2020.}
    \label{fig:caribbean_2020_012_blockbootstrap_50_resamplings}
\end{figure}

This procedure is then repeated a sufficient number of times, e.g.\@ using 25 or 50 resamplings, as in Figure~\ref{fig:caribbean_2020_012_blockbootstrap_50_resamplings}. All results shown in the report to illustrate the method use 50 resamplings.
We then apply our approach described in Section~\ref{sec:interp_percentile_method} to each of the resampled datasets individually, each time fitting the tail. Note that the $N_{\rm threshold}$ percentile is chosen once and for all (e.g.\@ for each resampled dataset, use the 99.99$^{\rm th}$ percentile as being the threshold)\footnote{Notice that technically the actual value $x_{\rm threshold}$ will depend on the exact realisation of each resampled dataset.}.

\bigskip

\begin{figure}[h]
    \centerline{\includegraphics[width=19pc, angle=0, trim = 0 0 0 0, clip]{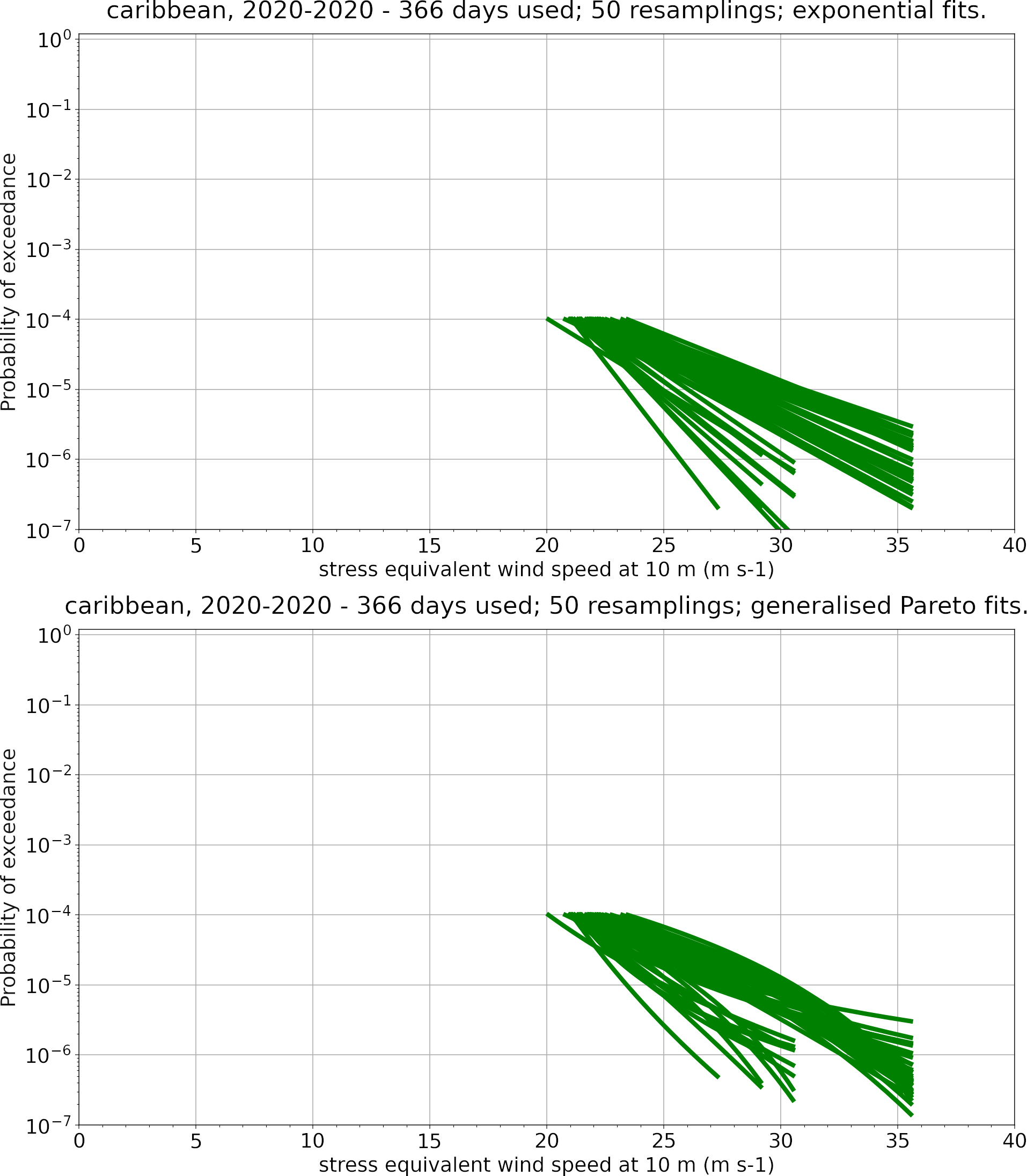}}
    \caption{Fitting individually each tail of the resampled datasets shown in Figure~\ref{fig:caribbean_2020_012_blockbootstrap_50_resamplings}, with exponential (\emph{top}) and generalised Pareto (\emph{bottom}). The threshold is chosen to be the 99.99$^{\rm th}$ percentile; i.e.\@ corresponding to a probability of exceedance of $10^{-4}$.}
    \label{fig:caribbean_2020_012_fits}
\end{figure}

After having fitted all the individual tails (cf.\@ Figure~\ref{fig:caribbean_2020_012_fits}), we then derive results for the considered basin, looking at the statistics: for each percentile value of interest (e.g.\@ the 99.999$^{\rm th}$ or the 99.9999$^{\rm th}$ percentiles), all the resampled datasets are used to finally calculate the corresponding mean value and variance. Interestingly, two different approaches can in fact be followed here:
\begin{itemize}
    \item The first one is exactly what Section~\ref{sec:interp_percentile_method} described. For percentiles higher than $N_{\rm threshold}$, each individual fit is used to replace the tail data; below the threshold however, the data in each resampled dataset itself is trusted to calculate percentiles directly. Results obtained with this approach will be referred to as ``fit''.
    \item Alternatively, having new samples at hand thanks to the block bootstrap method, one might instead decide to calculate a mean and a variance using directly the resampled data: simply calculating the mean and variance from the corresponding empirical percentiles of each individual resampled dataset. Results obtained with this approach, which does not make use of fits at all, but only of the block-bootstrap method, will be referred to as ``raw'' (seeing that this then does no use any modelling for the tail).
\end{itemize}
This gives us three different cases that we can consider: two using fits to stabilise the tail (exponential and generalised Pareto), and a coarser one using empirical percentiles calculated from the resampled tail datasets directly.\footnote{%
As expected, the latter, more crude, starts deviating at the lowest levels, due to the data being overly scarce.
}

\section{Sample results at basin level}

\begin{figure}[h!!!]
    \centerline{\includegraphics[width=19pc, angle=0, trim = 0 0 0 0, clip]{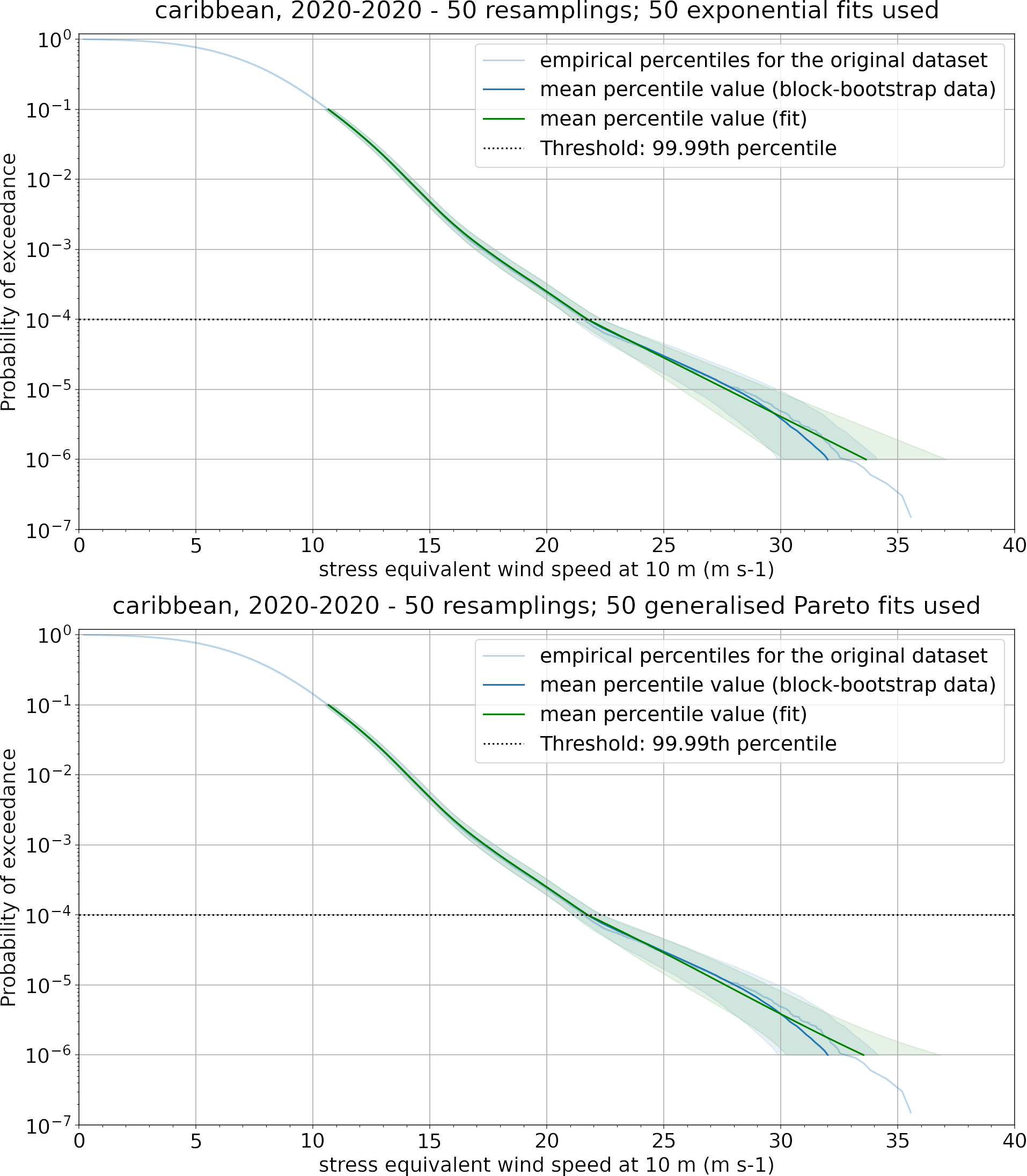}}
    \caption{ASCAT-A 012 data. Results using  exponential fits (\emph{top}) and generalised Pareto fits (\emph{bottom}).}
    \label{fig:caribbean_2020_012}
\end{figure}

\begin{figure*}[p]
    \includegraphics[width=\textwidth, trim = 0 0 0 0, clip]{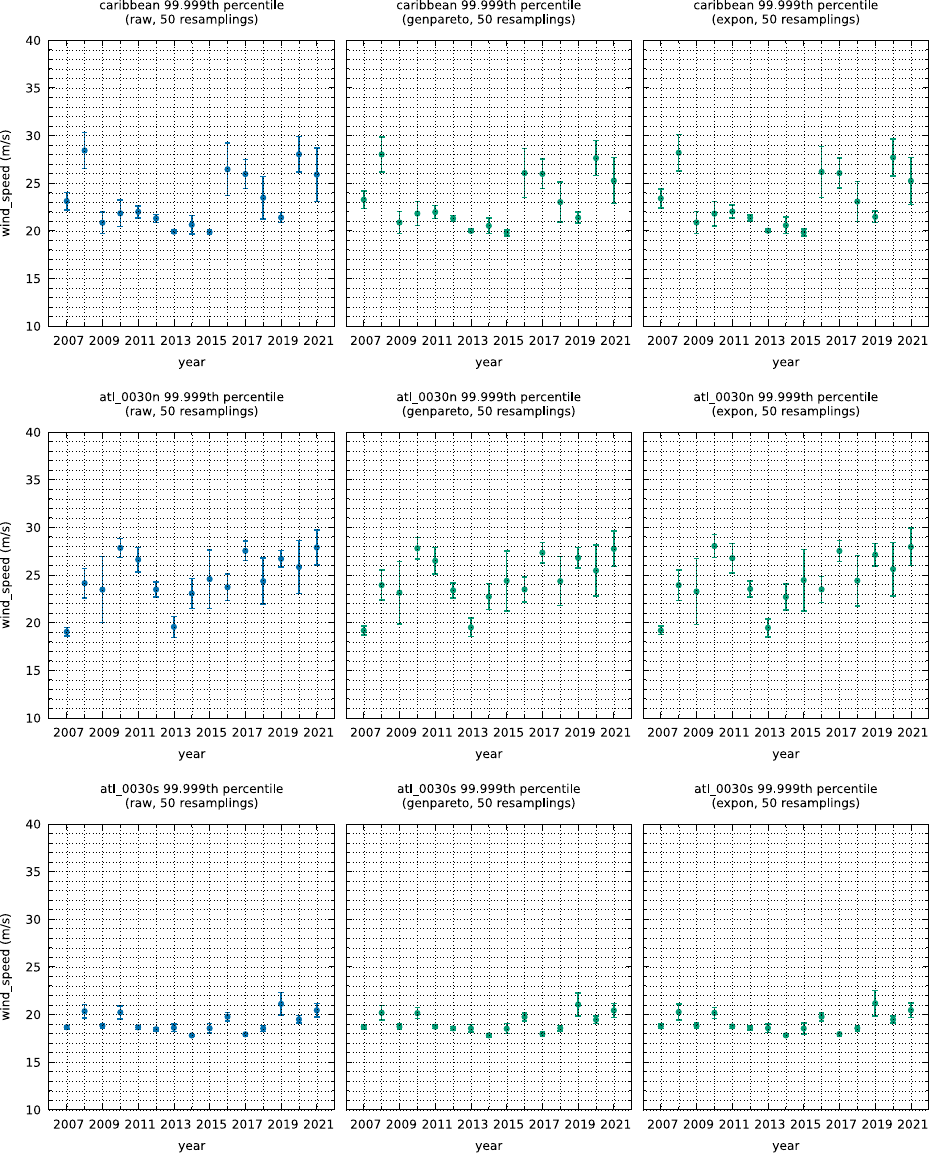}
    
    
    \caption{99.999$^{\rm th
    }$-percentile results for the 3 tropical basins, when using ASCAT-A 0.125°.}
    \label{fig:years__wind_speed_1e-5}
\end{figure*}

\begin{figure*}[p]
    \includegraphics[width=\textwidth, trim = 0 0 0 0, clip]{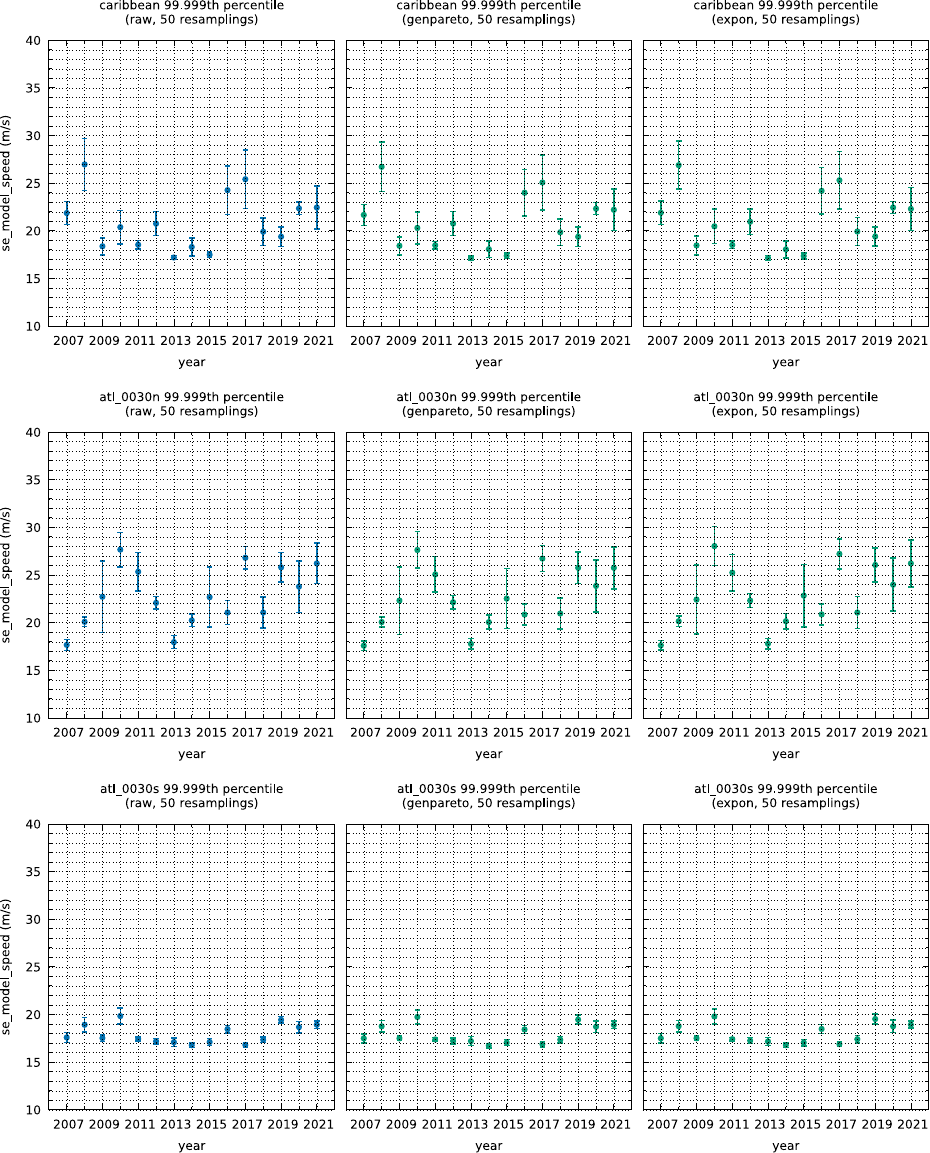}

    
    \caption{Same as Figure~\ref{fig:years__wind_speed_1e-5}, but for collocated ERA5 model winds.}
    \label{fig:years__se_model_speed_1e-5}
\end{figure*}

\begin{figure*}[p]
    \includegraphics[width=\textwidth, trim = 0 0 0 0, clip]{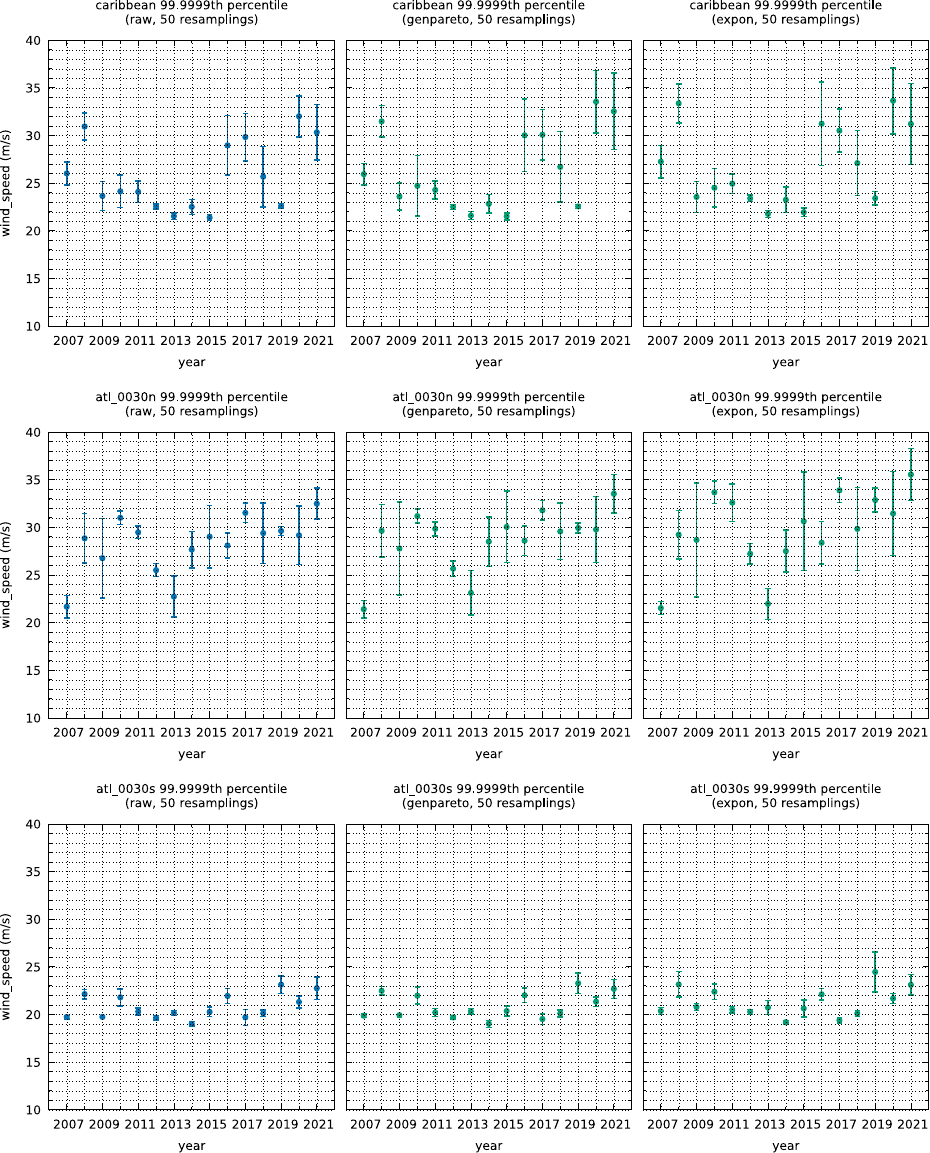}
    
    
    \caption{99.9999$^{\rm th
    }$-percentile results for the 3 tropical basins, when using ASCAT-A 0.125°.}
    \label{fig:years__wind_speed_1e-6}
\end{figure*}

As this section will demonstrate, the results obtained are very consistent. It may in fact be difficult to distinguish them down to a probability of exceedance of $10^{-5}$ (99.999$^{\rm th}$ percentile); not only in a specific case as in Figure~\ref{fig:caribbean_2020_012}, but even in general, cf.\@ in particular Figures~\ref{fig:years__wind_speed_1e-5} and~\ref{fig:years__se_model_speed_1e-5}. They moreover still remain close, even down to $10^{-6}$ as shown in Figure~\ref{fig:years__wind_speed_1e-6}.

\bigskip

Results have been obtained for every year from 2007 until 2021, for the Caribbean and the two Atlantic basins (0--30°~N/S), using ASCAT-A L3 data, both 012 and 025, using both the scatterometer 10-m stress-equivalent winds and the collocated ERA5 model winds. Here, we focus on results for the Caribbean basin using the year 2020; similar sample results for the Atlantic basins for the same year are given in the Appendix. Summary plots showing various results obtained using 012 data from 2007 until 2021, for all these three basins, in all cases, and for both scatterometer and model winds, are given in Figures~\ref{fig:years__wind_speed_1e-5} to~\ref{fig:years__wind_speed_1e-6}. 

\begin{figure}[h]
    \centerline{\includegraphics[width=19pc, angle=0, trim = 0 0 0 0, clip]{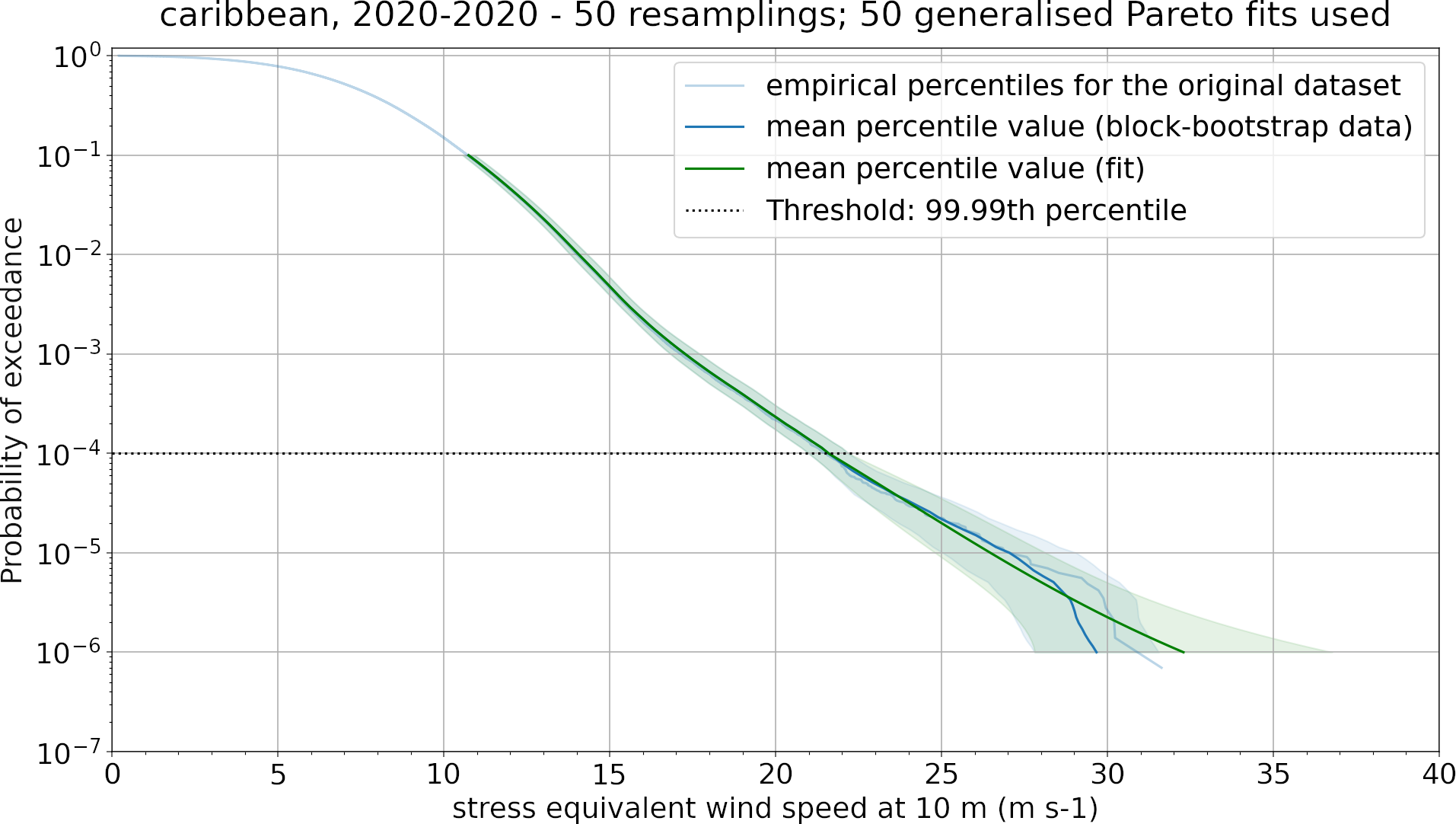}}
    \caption{Same as Figure~\ref{fig:caribbean_2020_012}, but using the corresponding 025 data.}
    \label{fig:caribbean_2020_025}
\end{figure}

\begin{figure}[h]
    \centerline{\includegraphics[width=19pc, angle=0, trim = 0 0 0 0, clip]{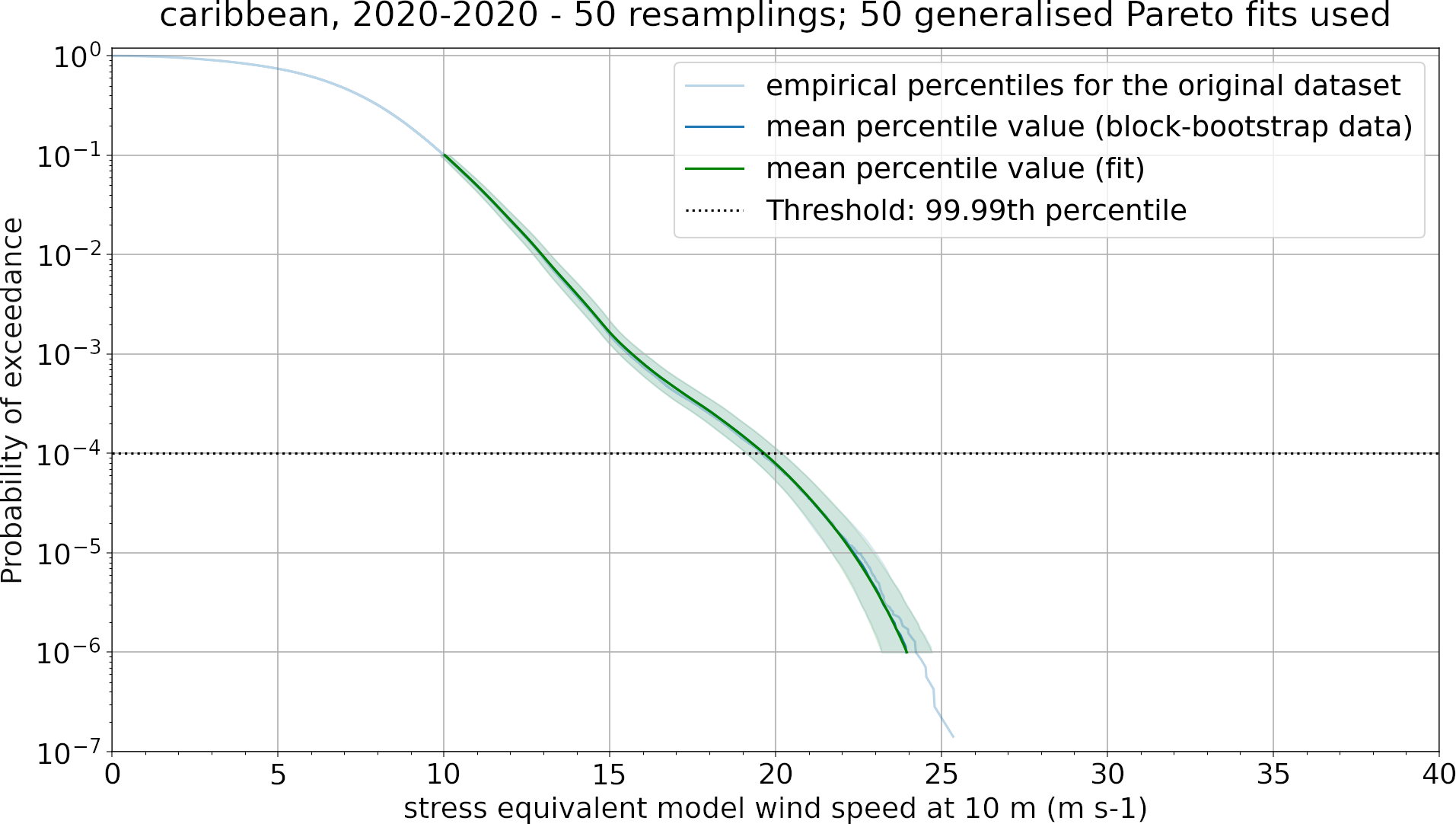}}
    \caption{Same as Figure~\ref{fig:caribbean_2020_012}, but using the 012 collocated ERA5 model wind speeds. Notice that ERA5 extreme winds are quite significantly lower.}
    \label{fig:caribbean_2020_012_model_wind_speeds}
\end{figure}

Note that the final results in this section are for the same Caribbean 2020 case used so far for illustration. These are shown in terms of the empirical probability of exceedance (for the one-to-one correspondence with percentiles, cf.\@ Eq.~\eqref{eq:link_exceedance_percentiles}).
First of all, for reference, we show in light blue the empirical probability of exceedance, calculated from the original ASCAT-A L3 dataset.
Then, the mean percentiles with estimated variance for ``fit'' and ``raw'' results are shown, in green and blue respectively. We stress one last time that each resampled dataset is fitted individually and that what is shown here as the final result is after calculating the mean wind-speed value and corresponding variance for each percentile.\footnote{In particular, the procedure is not that the ``raw'' blue curve would have to be fitted, nor the original dataset.}

\bigskip

Here, we show the following cases:
\begin{itemize}
    \item Figure~\ref{fig:caribbean_2020_012} shows the 012 results with ASCAT winds;
    \item Figure~\ref{fig:caribbean_2020_025}, the 025 results with ASCAT winds;
    \item Figure~\ref{fig:caribbean_2020_012_model_wind_speeds}, results using the 012 ERA5 model winds.
\end{itemize}

Comparing the exponential and generalised Pareto results in the \emph{top} and \emph{bottom} panels of Figure~\ref{fig:caribbean_2020_012}, it is clear that the obtained percentiles are very consistent in this case, and the agreement with the simpler ``raw'' method is also quite evident, even below a probability of exceedance of $10^{-5}$. Note that, in general, the generalised Pareto, having more parameters, is more flexible and can therefore lead to better fits of the tails.
A case showing a larger difference between exponential and generalised Pareto fits is given in the Appendix. The summary Figures~\ref{fig:years__wind_speed_1e-5} to~\ref{fig:years__wind_speed_1e-6} also allow one to compare results obtained with either exponential or generalised Pareto fits.

Results with both the coarser 025 scatterometer winds and the 012 ERA5 model winds lead to lower wind speeds, as expected; the 025 scatterometer winds being in quite good agreement with the higher resolution one (but of course with less data). We do notice that the ERA5 model winds are systematically quite significantly lower than the observations; see e.g.\@ Figures~\ref{fig:caribbean_2020_012_model_wind_speeds} and \ref{fig:years__se_model_speed_1e-5}.
For that reason, for comparing the different past instruments against ERA5, in the future we are also planning to use CDF-matching techniques (quantile mapping) as used in the OSI-SAF: the idea then being to create from ERA5 would-be ASCAT-A data, would-be QuikSCAT data and would-be ERS data over the last 30 years.

\section{Discussion and summary}

We find that the approach presented here leads to very robust results at basin level:
\begin{itemize}
    \item Robust against the use of either generalised Pareto fits or exponential fits of the tail, or even when avoiding the use of any fits at all and instead using empirical mean and standard deviations calculated from the resampled data obtained with the block-bootstrap method described here.
    \item The results obtained also behave as expected with decreasing resolution: 012 gives the highest winds, 025, slightly lower ones, and the coarser ERA5, even lower ones.
    \item There is enough data to use the 99.99$^{\rm th}$ percentile as threshold at basin level, leading to very consistent results with both 012 and 025 data (allowing for resolution differences): indeed, the results are extremely robust and stable down to a probability of exceedance of $10^{-5}$ (99.999$^{\rm th}$ percentile)\hspace{1pt}---\hspace{1pt}which is our main result\hspace{1pt}---\hspace{1pt}and still quite consistent within the uncertainties even slightly lower than that, such that the method seems trustworthy at least down to the 99.9999$^{\rm th}$ percentile (at which point slight differences can start to appear, depending on how the tail is being modelled).
\end{itemize}

We moreover checked that:

\begin{itemize}
    \item The results appear robust against changes in the number of resampling (with the 99.9999$^{\rm th}$ percentile for the atl\_0030n basin with 025 data in 2020 staying within about merely 1~m/s, when the number of resamplings is anywhere between 25 and 150).
    \item With a careful/conservative stance (neither extrapolating, nor trying to go too low) and notably focusing on the predicted 99.999$^{\rm th}$ percentile, the results are actually also quite robust when changing the threshold by a factor 10 up or down in probability of exceedance. The 99.99$^{\rm th}$ percentile generally appears to be the best choice for the amount of data at hand. There really is a good agreement of results obtained from either fits or simply empirically based on the block-bootstrap resampled data down to $10^{-5}$ and even $10^{-6}$. Using a lower percentile as threshold might not capture well enough the lower part of the tail (trusting the shape of the model more than the tail) and a larger one might lead to overfitting (overly trusting scarce data). There is no automatic way to determine the best threshold; one must look at the data and results. 
    When the data is becoming scarce, the question of what we trust more becomes more pressing. We believe that we should avoid having to make our results depend on a strong assumption about what to trust; this can be achieved by being more conservative. After all, we do not believe that decadal trends would only appear at the level of the 10$^{-7}$ probability of exceedance and not already at 10$^{-6}$ or 10$^{-5}$.
\end{itemize}

Given its stability and the fact that it allows peering at extreme percentiles that are high enough to correspond to tropical cyclone winds (the 99$^{\rm th}$ percentile, in comparison, can be too low, i.e.\@ below the change of curvature seen in the probability of exceedance), the method therefore appears to be a powerful tool to assess the existence of trends in the wind speeds of extreme storms.

As already stressed in the introduction, this method should actually be applied to earlier instruments as well, to extend the time range, before we can draw conclusions about the existence of decadal trends (also to avoid being overly sensitive to variations of e.g.\@ the ENSO index). 
Our conclusion remains that a longer period is required before one can hope to reach a sensible conclusion on decadal trends that would not solely rest upon a decade of data.

On this aspect, \citet{Lee_etal:2020} explicitly calls for climate-quality time series of wind speeds. For scatterometry, these may be obtained in an overlapping series from the ERS scatterometer (1991--1999), QuikSCAT Seawinds (1999--2009) and ASCAT (2007--today). The stability of these instruments is proven \citep{stoffelen2021} and may together provide the basis for a long stable data series. In addition, this will be continued with the successor of ASCAT on board of MetOp Second Generation: SCA.

\acknowledgments

The work of A.\@ P.\@ and that of A.\@ S.\@ were supported by the ESA MAXSS (Marine Atmosphere eXtreme Satellite Synergy) project of the European Space Agency, under the contact 4000132954/20/I-NB.
The authors declare no conflict of interest.

%
%
\datastatement

This study has been conducted using E.U.\@ Copernicus Marine Service Information (https://doi.org/10.48670/moi-00183).

%

\appendix





%



\section*{Sample results for the two Atlantic basins (0--30°~N/S)}

\begin{figure}[h]
    \centerline{\includegraphics[width=19pc, angle=0, trim = 0 0 0 0, clip]{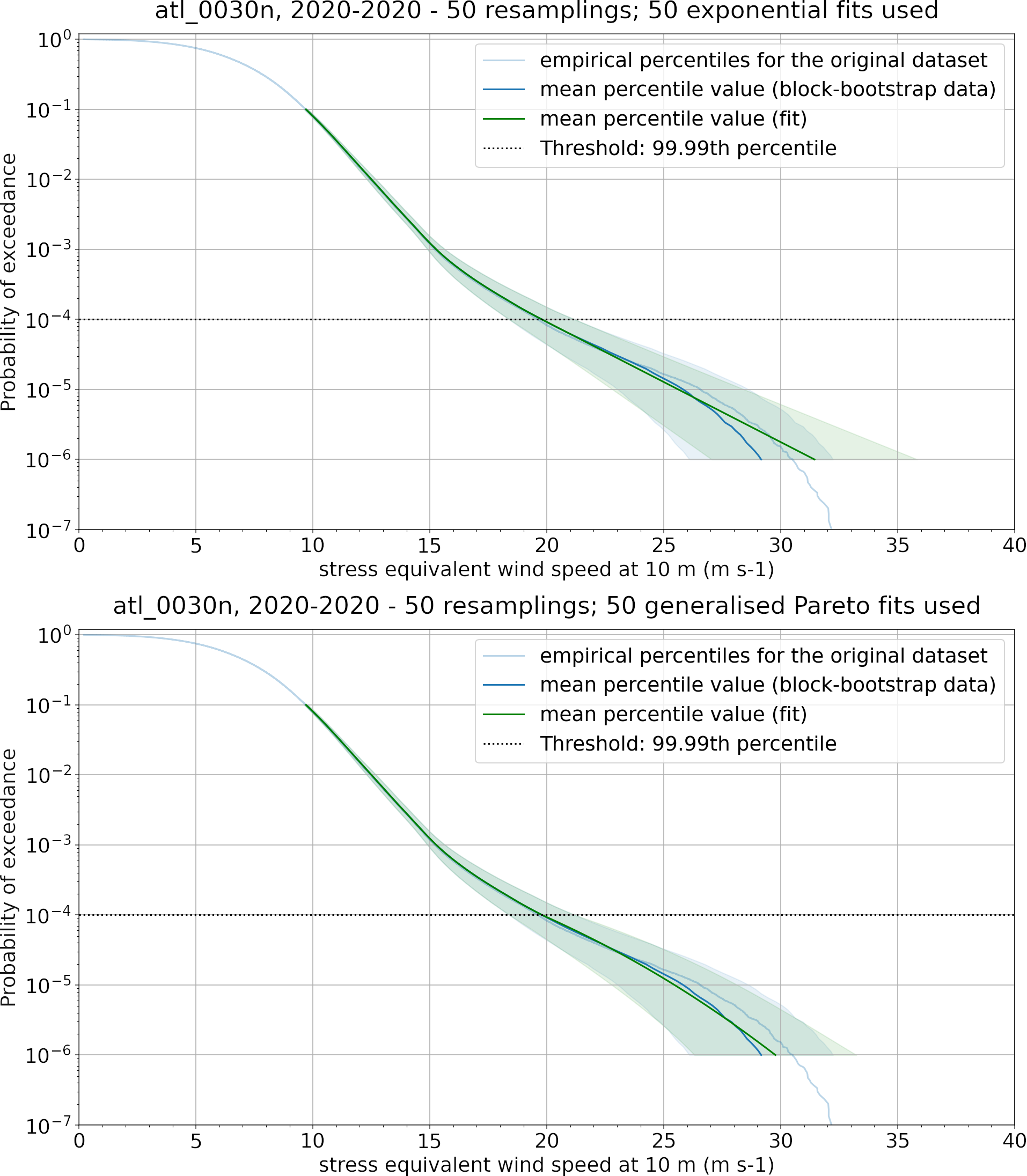}}
    \caption{Same as Figure~\ref{fig:caribbean_2020_012}, but for the North Atlantic basin (atl\_0030n).}
    \label{fig:atl_0030n_2020_012}
\end{figure}

\begin{figure}[h]
    \centerline{\includegraphics[width=19pc, angle=0, trim = 0 0 0 0, clip]{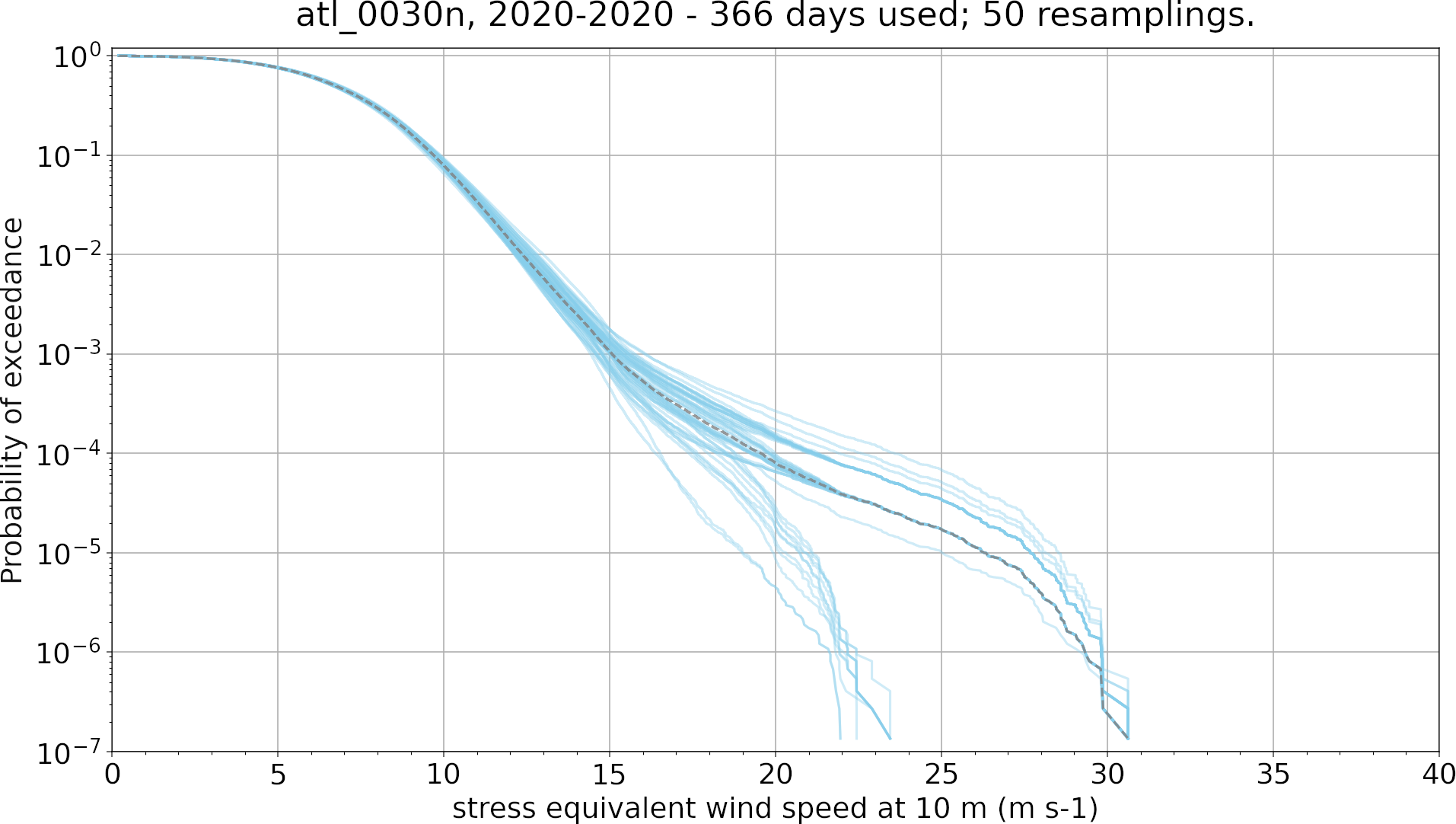}}
    \caption{ASCAT-A 025 data. Empirical probability of exceedance calculated for 50 block-bootstrap resamplings of the North Atlantic basin (0--30°~N) wind speeds, for the year 2020.}
    \label{fig:atl_0030n_2020_012_blockbootstrap_50_resamplings}
\end{figure}

\begin{figure}[h]
    \centerline{\includegraphics[width=19pc, angle=0, trim = 0 0 0 0, clip]{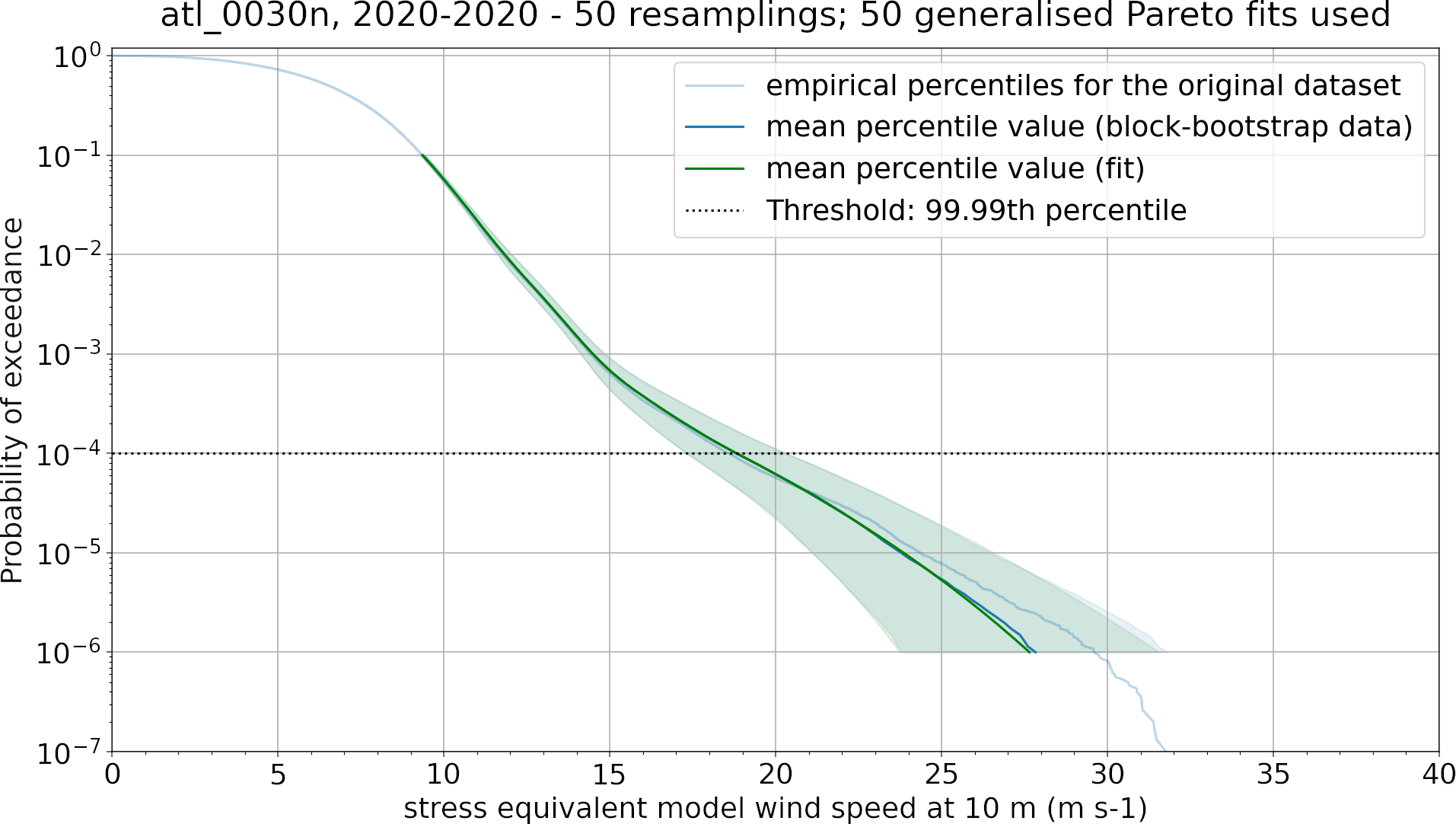}}
    \caption{Same as Figure~\ref{fig:atl_0030n_2020_012}, but for collocated ERA5 model wind speeds using generalised Pareto.}
    \label{fig:atl_0030n_2020_012_model_wind_speeds}
\end{figure}

\begin{figure}[h]
    \centerline{\includegraphics[width=19pc, angle=0, trim = 0 0 0 0, clip]{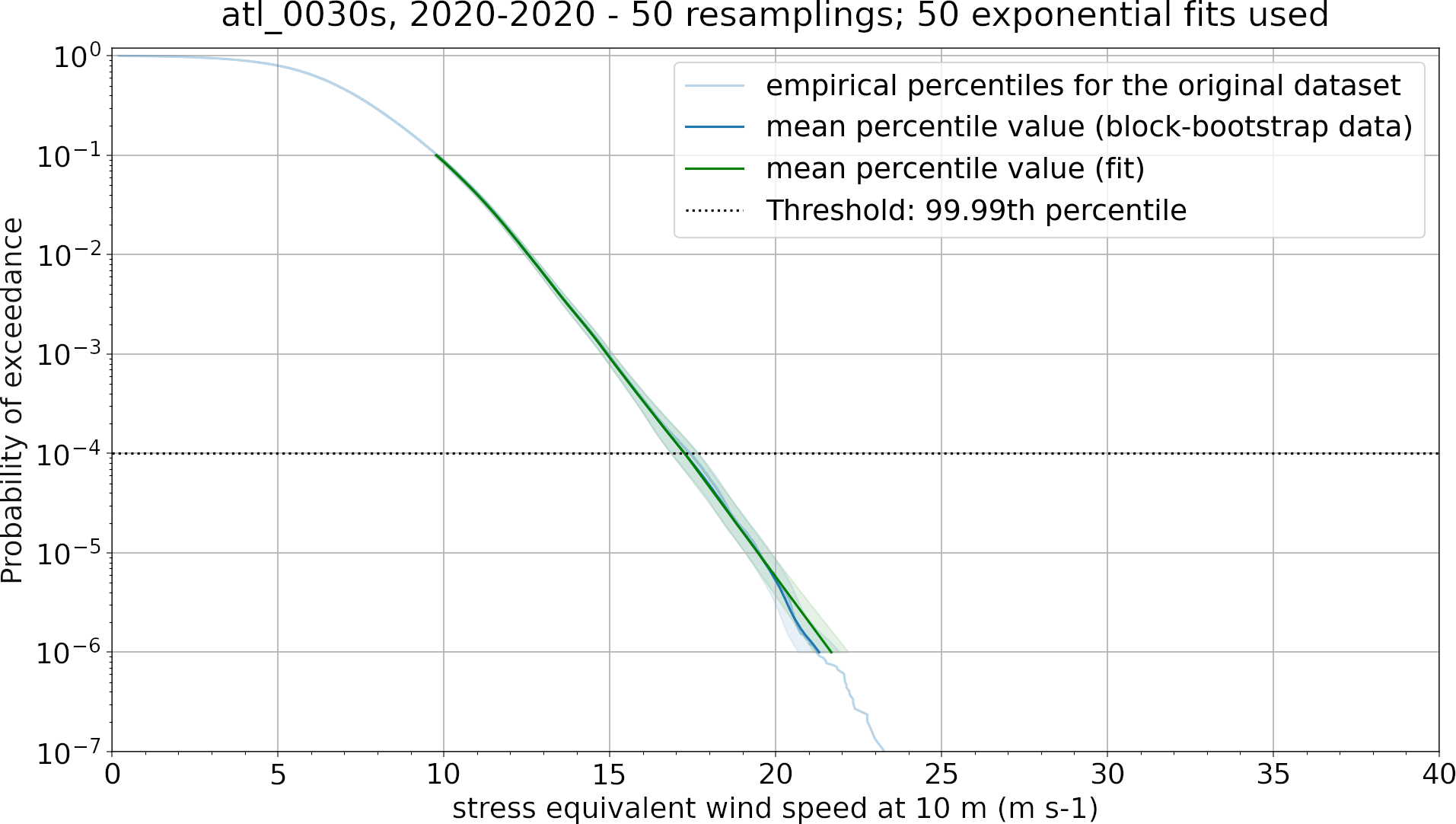}}
    \caption{Same as Figure~\ref{fig:atl_0030n_2020_012}, but this time for the South Atlantic basin (0--30°~S). Notice how there is no change in curvature in the probability of exceedance where TCs would be expected and that the tail is essentially exponential\hspace{1pt}---\hspace{1pt}as can be expected as there are essentially no TCs in the South Atlantic.}
    \label{fig:atl_0030s_2020_012}
\end{figure}

\begin{figure}[h]
    \centerline{\includegraphics[width=19pc, angle=0, trim = 0 0 0 0, clip]{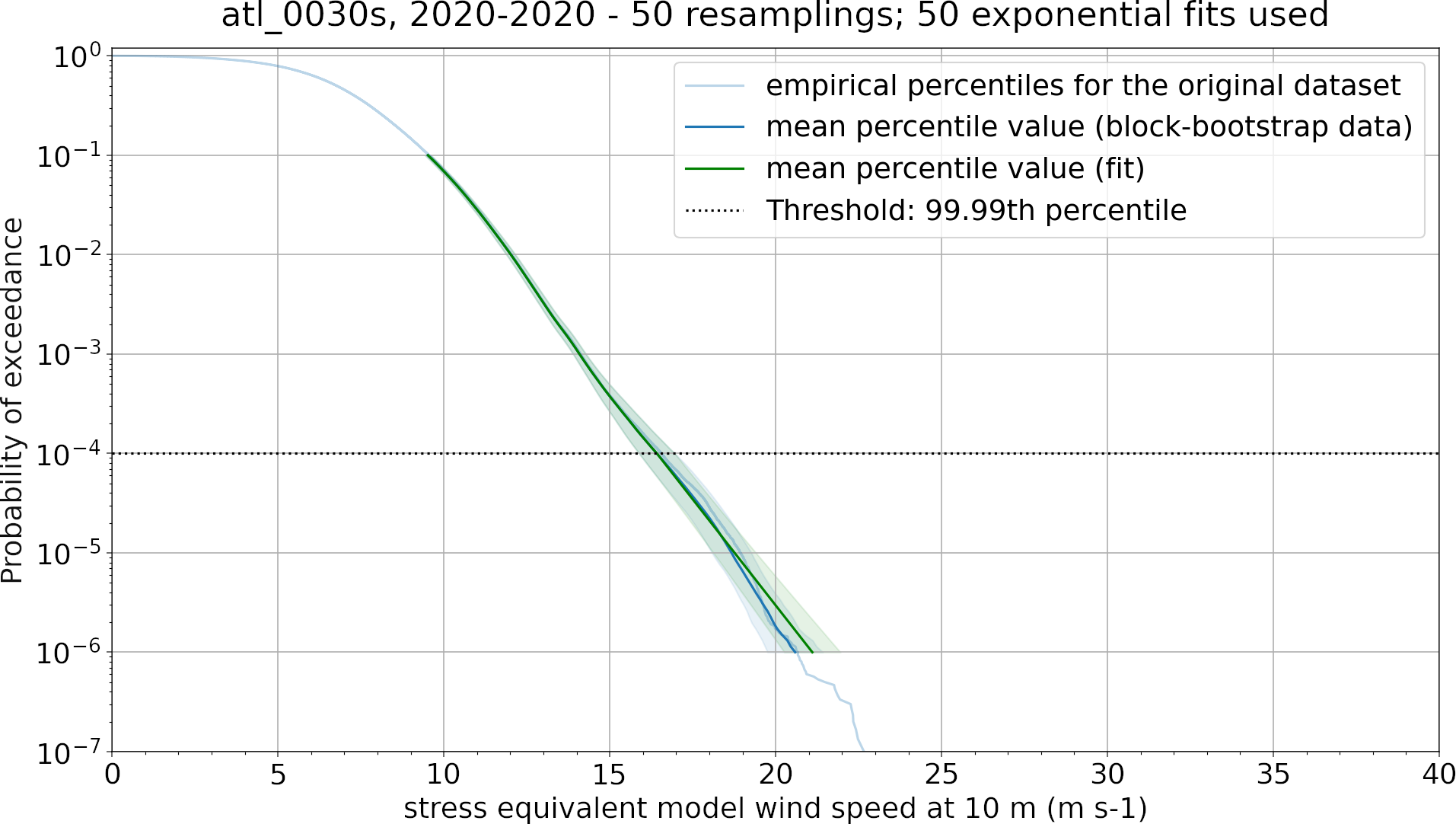}}
    \caption{Same as Figure~\ref{fig:atl_0030s_2020_012}, but for collocated ERA5 model wind speeds using exponential fits.}
    \label{fig:atl_0030s_2020_012_model_wind_speeds}
\end{figure}

This section provides sample plots similar to those discussed in the main text, but in the case of the two other tropical basins considered, also for the year 2020: i.e.\@ the North Atlantic basin up to 30°N (Figures~\ref{fig:atl_0030n_2020_012} to~\ref{fig:atl_0030n_2020_012_model_wind_speeds}) and the South Atlantic basin down to 30°S (Figures \ref{fig:atl_0030s_2020_012} and \ref{fig:atl_0030s_2020_012_model_wind_speeds}).

\bibliographystyle{ametsocV6}
\bibliography{references}

\end{document}